\documentclass[12pt]{iopart}
\usepackage{iopams}  
\usepackage[]{graphicx}

\usepackage{epsfig}
\usepackage{wrapfig}
\usepackage{color}

\def\##1{{\bf #1}}
\def\=#1{\underline{\underline #1}}

\def\.{\mbox{ \tiny{$^\bullet$} }}

\def\les{\left[}
\def\ris{\right]}
\def\lec{\left\{}
\def\ric{\right\}}

\def\eps{\varepsilon}
\def\epso{\eps_{\scriptscriptstyle 0}}
\def\lambdao{\lambda_{\scriptscriptstyle 0}}
\def\muo{\mu_{\scriptscriptstyle 0}}
\def\ko{k_{\scriptscriptstyle 0}}
\def\etao{\eta_{\scriptscriptstyle 0}}

\def\epsa{\eps_a}
\def\epsb{\eps_b}
\def\epsc{\eps_c}
\def\epsdt{\eps_d}
\def\epsmet{\eps_{met}}
\def\chiv{\chi_v}

\def\lmet{L_{met}}

\def\ux{\hat{\#u}_x}
\def\uy{\hat{\#u}_y}
\def\uz{\hat{\#u}_z}

\def\tildeQ{{\tilde Q}}

\begin{document}

\title[SPP waves guided by a metal slab in an SNTF]{Surface-plasmon-polariton wave propagation guided by a metal slab in a sculptured nematic thin film}

\author{Muhammad Faryad and Akhlesh Lakhtakia\footnote{Corresponding author}}

\address{Nanoengineered Metamaterials Group (NanoMM), Department of Engineering Science and
Mechanics, 
Pennsylvania State University, University Park, PA  16802-6812, USA}
\ead{akhlesh@psu.edu}

\begin{abstract}

Surface-plasmon-polariton~(SPP) wave propagation guided by a metal slab in a periodically nonhomogeneous sculptured nematic thin film~(SNTF) was studied theoretically. The morphologically significant planes of the SNTF on both sides of the metal slab could either be aligned or twisted with respect to each other. The canonical boundary-value problem was formulated, solved for SPP-wave propagation, and examined to determine the effect of slab thickness on the multiplicity and the spatial profiles of SPP waves. Decrease in slab thickness was found to result in more intense coupling of two metal/SNTF interfaces. But when the metal slab becomes thicker, the coupling between the two interfaces
reduces and  SPP waves localize to one of the two  interfaces. The greater the coupling between the two metal/SNTF
interfaces, the smaller is the phase speed.

\vspace{3 mm}

\end{abstract}
\maketitle

\newpage
\section{Introduction}

Quasi-particles formed collectively by plasmons and polaritons bound to the interface of a metal and a dielectric material are called
surface plasmon-polaritons (SPPs)~\cite{Zayats,Mbook,Felbacq,PLreview}. Although SPPs are quantum objects,
trains of SPPs traveling along the interface are
classically viewed as SPP waves. Much recent literature is devoted to the exploitation of SPP waves for diverse technological
purposes, including sensing,   imaging, and communications \cite{Hbook,Kalele,AZLem,DD2008}.

This technoscientific ferment is due to a resonance phenomenon that arises
when the energy carried by photons in the dielectric material is transferred to  free electrons in the metal at that interface,
and vice versa. Different dielectric materials
will become differently polarized on interrogation by an electromagnetic field, thereby enabling a widely used
technique for sensing chemicals and biochemicals \cite{Hbook,AZLem}. Furthermore,
SPP imaging systems are used for high-throughput analysis of biomolecular interactions---for
proteomics, drug discovery, and pathway elucidation  \cite{AZLchapter, Aoki}.
SPP-based imaging technology
has been successfully applied to the screening of bioaffinity interactions with DNA, carbohydrates,
peptides, phage display libraries, and proteins \cite{Kanda}. SPP-based imaging techniques are
also going to be useful for   lithography \cite{Mbook,KMA}. Finally,
as SPP waves can be excited in the terahertz and optical  regimes, they
may be useful for high-speed information on computer chips \cite{MBKMRA}. Whereas conventional
wires are very attenuative at frequencies beyond a few tens of GHz, ohmic losses are minimal for
plasmonic transmission \cite{DD2008} which enables long-range communications \cite{Berini}.

Even a cursory perusal of literature will show that only one SPP wave at a specific frequency is excited along the interface,
if the partnering dielectric material is  homogeneous. 
But if this material were to be periodically nonhomogeneous in
the direction normal to the interface, the number of SPP waves---of different phase speeds, attenuation rates, and field structures,
but the same frequency---can be higher. The material can be   isotropic \cite{YJJ2009}
or anisotropic  \cite{PLprsa,PartII}, the latter obviously providing more options due to birefringence \cite{MLbook}.
The possibility of exciting multiple SPP waves provides exciting prospects for enhancing the scope of the applications of SPP waves.
For sensing applications,
the use of more than one distinct SPP waves would increase confidence in a reported measurement; also,
more than one analyte could be sensed at the same time, thereby increasing the capabilities
of multi-analyte sensors.    For imaging applications, the simultaneous creation of two images
may become possible. For plasmonic communications, the availability of multiple channels
would make information transmission more reliable as well as enhance capacity.  

These possibilities are promised by the use of
sculptured thin films~(STFs) as the partnering dielectric material.
STFs are anisotropic and unidirectionally nonhomogeneous thin films whose permittivity profile has been nanoengineered during the 
physical vapor deposition process~\cite{Lmsec,LMbook}.  The possibility of multiple SPP waves
localized to the single interface of a metal and a periodically
nonhomogeneous
sculptured nematic thin film (SNTF) has been demonstrated
both theoretically \cite{PartII,PartI,PartIV} and
experimentally \cite{PartIII}. An SNTF is a special type of an STF---with a permittivity profile which is sculptured, during physical
vapor deposition, only in one plane
called the morphologically significant plane \cite{LMbook}. Also, SPP wave propagation along the interface of a metal and a chiral STF has also been shown
both theoretically~\cite{PL2009} and experimentally~\cite{DPL} to admit more than one type of SPP waves. 

A further increase in the number of SPP waves would require the use of multiple parallel metal/dielectric interfaces,
which is already established well with   isotropic dielectric materials \cite{Economou,Wendler,Yang}. Pursuing this line of thinking,
 we decided to solve the canonical problem of wave propagation localized to a sufficiently thin metallic slab
inserted in an SNTF.  As our supposition was confirmed in  a preliminary study \cite{FL2010}, we conducted a full-scale
investigation, the details and the results of which are reported here.
In Sec.~\ref{theory}, we present the theoretical formulation of the canonical boundary-value problem, where a dispersion equation is obtained. 
Representative numerical results are discussed in Sec.~\ref{nrd}, and concluding remarks are given in Sec.~\ref{conc}. 

An $\exp(-i\omega t)$ time-dependence is implicit, with $\omega$
denoting the angular frequency, $t$ the time, and $i=\sqrt{-1}$. The free-space wavenumber, the
free-space wavelength, and the intrinsic impedance of free space are denoted by $\ko=\omega\sqrt{\epso\muo}$,
$\lambdao=2\pi/\ko$, and
$\etao=\sqrt{\muo/\epso}$, respectively, with $\muo$ and $\epso$ being  the permeability and permittivity of
free space. Vectors are in boldface, dyadics are underlined twice,
column vectors are in boldface and enclosed within square brackets, and
matrixes are underlined twice and square-bracketed. The asterisk denotes the complex conjugate,
and the Cartesian unit vectors are
identified as $\ux$, $\uy$, and $\uz$.

\section{Canonical Boundary-Value Problem}\label{theory}

Suppose that the region $L_-\leq z\leq L_+$ is occupied by an isotropic and homogeneous metal
with
complex-valued relative permittivity scalar $\epsmet$. The thickness of the metal slab is denoted
by $\lmet=L_+-L_-$.

The regions $z\gtrless L_\pm$ are occupied by the chosen SNTF with periodically nonhomogeneous  permittivity dyadic \cite{PartII,PartI}
\begin{equation}
\label{epssntf}
\=\eps_{SNTF}(z)= \epso\, \=S_z(\gamma^\pm)\cdot
\=S_{y}(z) \cdot \={\epso}_{ref}(z)\cdot \=S_{y}^{-1}(z)\cdot\=S_z^{-1}(\gamma^\pm)
\,,\quad z\gtrless L_{\pm}\,,
\end{equation}
where the locally orthorhombic symmetry is expressed through the diagonal
dyadic
\begin{equation}
\={\epso}_{ref}(z)=\epsa(z)\,\uz\uz+\epsb(z)\,\ux\ux+\epsc(z)\,\uy\uy
\end{equation}
and the local tilt dyadic
\begin{equation}
\=S_{y}(z) =(\ux\ux+\uz\uz) \cos\left[\chi(z)\right]  +
(\uz\ux - \ux\uz) \sin \left[\chi(z)\right] +\uy\uy
\end{equation}
expresses nematicity.
 Both the relative permittivity scalars $\eps_{a,b,c}(z)$ and the tilt angle $\chi(z)$
are supposed to have been nano-engineered by a periodic variation of  the direction of the vapor flux during
fabrication by physical vapor deposition \cite{LMbook,PartIII}. This periodic variation is captured by
 the vapor incidence angle \cite{PartIII}
\begin{equation}
\chiv (z) = {\tilde\chi}_v \pm \delta_v\,\sin\left[\frac{\pi (z-L_\pm)}{\Omega}\right]\,,\quad z\gtrless L_{\pm}\,,
\end{equation}
that varies sinusoidally with $z$.  The third dyadic in Eq.~(\ref{epssntf}) was chosen as
\begin{equation}
\=S_z(\gamma^\pm) = (\ux\ux+\uy\uy) \cos\gamma^\pm + (\uy\ux-\ux\uy) \sin\gamma^\pm
+\uz\uz\,,
\end{equation}
so that plane formed by the unit vectors $\uz$ and $\ux\cos\gamma^\pm+\uy\sin\gamma^\pm$
is the morphologically significant plane for $z\gtrless L_\pm$. Thus, there is
 sufficient flexibility in the formulation with respect to the twist $\gamma^+-\gamma^-$ of the 
 two morphologically significant planes.
Note that $\gamma^+=\gamma^-$ in our preliminary
 report \cite{FL2010}, but that restriction was removed for the full-scale investigations reported here.

Without loss of generality, let us choose the direction of SPP-wave propagation  
in the $xy$ plane to be parallel to the $x$ axis.  Accordingly, we set
\begin{equation}
\left.
\begin{array}{c}
\#E(\#r)=\#e(z)\,\exp\left( i\kappa x\right)\\
\#H(\#r)=\#h(z)\,\exp\left( i\kappa x\right)
\end{array}\right\}
\,,
\end{equation}
where $\kappa$ is a complex-valued scalar. 

The axial
field components $e_z(z)$ and $h_z(z)$ can be expressed in terms of the column vector
\begin{equation}
\left[\#f(z)\right]= \left[e_x(z)\quad e_y(z) \quad h_x(z)\quad h_y(z)\right]^T\,
\end{equation}
via
\begin{equation}
\kappa
\les\begin{array}{c}
e_z(z)\\
0\\
h_z(z)\\
0
\end{array}\ris= \left\{
\begin{array}{ll}
\left[\=A^\pm(z)\right]\cdot\left[\#f(z)\right], \quad &
z\gtrless L_{\pm}\,,
\\[4pt]
 \left[\=A^{met}(z)\right]\cdot\left[\#f(z)\right], \quad &
z\in\left(L_-,L_+\right)\,,
\end{array}
\right.
\end{equation} 
where one   4$\times$4 matrix
\begin{eqnarray}
\nonumber
&&\left[\=A^\pm(z)\right]= \les\begin{array}{cccc}
0 & 0 & 0 & -\frac{\kappa^2}{\omega\epso}\,
\frac{\epsdt(z)}{\epsa(z)\,\epsb(z)}\\
0 & 0 & 0& 0 \\
0 & \frac{\kappa^2}{\omega\muo} & 0 & 0\\
0 & 0 & 0 & 0
\end{array}\ris
\\[4pt]
&&  \hspace{-2cm}
+\,\kappa\,\frac{\epsdt(z)\,\les\epsa(z)-\epsb(z)\ris}{\epsa(z)\,\epsb(z)}\,
\sin\les\chi(z)\ris\,\cos\les\chi(z)\ris\,
\les\begin{array}{cccc}
\cos\gamma^\pm& \sin\gamma^\pm & 0 & 0\\
0& 0 & 0 & 0\\
0 & 0 & 0 & -\sin\gamma^\pm\\
0 & 0 & 0& \cos\gamma^\pm
\end{array}\ris\,
\end{eqnarray}
involves the auxiliary quantity
\begin{equation}
\epsdt(z)=  \epsa(z)\,\epsb(z)/
\lec \epsa(z)\,\cos^2\les\chi(z)\ris + \epsb(z)\,\sin^2\les\chi(z)\ris\ric
\,,
\end{equation}
and the other  4$\times$4 matrix
\begin{equation}
\left[\=A^{met}(z)\right]=
\les\begin{array}{cccc}
0 & 0 & 0 & -\frac{\kappa^2}{\omega\epso\epsmet}\\
0 & 0 & 0& 0 \\
0 & \frac{\kappa^2}{\omega\muo} & 0 & 0\\
0 & 0 & 0 & 0
\end{array}\ris\,.
\end{equation}

The column vector $\left[\#f(z)\right]$ satisfies the matrix differential equations
\begin{equation}
\label{MODEsntf}
\frac{d}{dz}\left[\#f(z)\right]=i \left[\=P^\pm(z)\right]\cdot\left[\#f (z)\right]\,,
\quad z\gtrless L_\pm\,,
\end{equation}
and
\begin{equation}
\label{MODEmet}
\frac{d}{dz}\left[\#f(z)\right]=i \left[\=P^{met}(z)\right]\cdot\left[\#f (z)\right]\,,\quad
z\in\left(L_-,L_+\right)\,,
\end{equation}
where the 4$\times$4 matrixes
\begin{eqnarray}
\nonumber
&&\left[\=P^\pm(z)\right]= \left[\=A^\pm(z)\right]+
\\[4pt]
&&\hspace{-2.5cm}
\omega \les\begin{array}{cccc}
0 & 0 & 0 & \muo \\
0 & 0 & -\muo & 0 \\
\epso \left[\epsc(z)-\epsdt(z)\right]\cos\gamma^\pm\,\sin\gamma^\pm & 
-\epso\left[\epsc(z)\cos^2\gamma^\pm+\epsdt(z)\sin^2\gamma^\pm\right]
& 0 & 0\\
\epso\left[\epsc(z)\sin^2\gamma^\pm+\epsdt(z)\cos^2\gamma^\pm\right] & -\epso \left[\epsc(z)-\epsdt(z)\right]\cos\gamma^\pm\,\sin\gamma^\pm & 0 & 0
\end{array}\ris
\label{eq7.14}
\end{eqnarray}
and
\begin{equation}
\hspace{-2cm}
\left[\=P^{met}(z)\right]= \left[\=A^{met}(z)\right]+
\omega\les\begin{array}{cccc}
0 & 0 & 0 & \muo \\
0 & 0 & -\muo & 0 \\
0 & 
-\epso\epsmet
& 0 & 0\\
\epso\epsmet& 0 & 0 & 0
\end{array}\ris
\,.
\label{Pmet}
\end{equation}

Equation (\ref{MODEmet}) can be solved straightforwardly to yield
\begin{equation}
\left[\#f(L_+) \right]= \exp\lec i \les\=P^{met}\ris(L_+-L_-)\ric\cdot
\left[\#f(L_- \right]\,.
\label{meteq}
\end{equation}
Equation (\ref{MODEsntf}) requires numerical
solution    by the piecewise uniform approximation
technique \cite{PartIV,APL2009}. The ultimate aim is to determine the matrixes
$\left[  \={Q}^{\pm}\right]  $ that appear in the relations%
\begin{equation}
\left[\#f(L_\pm\pm  2\Omega)  \right]
=\left[  \={Q}^{\pm}\right]  \cdot 
\left[\#f(  L_\pm)  \right]
\end{equation}
to characterize the optical response of one period of the SNTF on either side
of the metal slab.
Basically, this technique consists of subdividing each period of the SNTF into
a cascade of electrically thin sublayers parallel to the plane $z=0$, and
assuming the dielectric properties to be spatially uniform in each sublayer. A
sufficiently large number $N+1$\ points $z_{n}^{\pm}=L_\pm\pm2\Omega\left(
n/N\right)  $, $n\in\lbrack0,N]$, are defined on each side of the metal slab
and the matrixes%
\begin{equation}
[\={W}_{n}^{\pm}]  =\exp\left\{  \pm i\left[
\={P}^{\pm}\left(  \frac{z_{n-1}^{\pm}+z_{n}^{\pm}}%
{2}\right)  \right]  \frac{2\Omega}{N}\right\} \, ,
\quad
n\in[1,N]\,,
\end{equation}
are calculated for a specific value of $\kappa$; then%
\begin{equation}
[  \underline{\underline{Q}}^{\pm}]  \cong[
\underline{\underline{W}}_{N}^{\pm}]  \cdot[
\underline{\underline{W}}_{N-1}^{\pm}]  \cdot...\cdot[
\underline{\underline{W}}_{2}^{\pm}]  \cdot[
\underline{\underline{W}}_{1}^{\pm}]\,  .  
\end{equation}
A sublayer thickness $2\Omega/N=2\ $nm was adequate for
the results reported in Sec.~\ref{nrd}.

By virtue of
the Floquet--Lyapunov theorem \cite{YS1975}, we can define the matrixes
$[  \={\tildeQ}^{\pm}]  $  such that%
\begin{equation}
[\={Q}^{\pm}]     =\exp\left\{\pm  i2\Omega
[\={\tildeQ}^{\pm}]  \right\}\,.  
\end{equation}
Both  $[\=Q^\pm]$ and $[\=\tildeQ^\pm]$ share the same eigenvectors, and
their eigenvalues are also related as follows. Let $\left[  \#{t}%
^{\pm}\right]  ^{\left(  n\right)  }$, $n\in[1,4] $, be the
eigenvector corresponding to the \textit{n}th eigenvalue $\sigma_{n}^{\pm}%
$\ of $\left[  \underline{\underline{Q}}^{\pm}\right]  $; then, the
corresponding eigenvalue $\alpha_{n}^{\pm}$\ of $[\=\tildeQ^{\pm}]  $ is given by%
\begin{equation}
 \alpha_{n}^{\pm}  =\mp i\frac{\ln\sigma_{n}^{\pm}}{2\Omega}\,.
\end{equation}

The electromagnetic fields of the SPP wave must diminish in
magnitude as $z\rightarrow\pm\infty$. Therefore, in the half-space $z>L_+$,
we first label the eigenvalues of $[  \underline{\underline{\tildeQ}}^{+}]$ such that Im$\left[  \alpha_{1,2}^{+}\right]  >0$ and then set%
\begin{equation}
\left[  \#f\left(  L_+\right)  \right]  =\left[
\begin{array}
[c]{cc}%
\left[  \#{t}^{+}\right]  ^{\left(  1\right)  } & \left[
\#{t}^{+}\right]  ^{\left(  2\right)  }%
\end{array}
\right]  \cdot\left[
\begin{array}
[c]{c}%
A^+_{1}\\[5pt]
A^+_{2}%
\end{array}
\right]  , \label{eq18}%
\end{equation}
where $A^+_{1,2}$  are unknown scalars; the other two eigenvalues of
$[ \underline{\underline{\tildeQ}}^{+}]  $ describe fields that
amplify as $z\rightarrow+\infty$\ and cannot therefore contribute to the
SPP wave. A similar argument for the half-space $z<L_-$ requires us to
ensure that Im$\left[  \alpha_{1,2}^{-}\right]  <0$ and then to set%
\begin{equation}
\left[  \#f\left(  L_-\right)  \right]  =\left[
\begin{array}
[c]{cc}%
\left[  \#{t}^{-}\right]  ^{\left(  1\right)  } & \left[
\#{t}^{-}\right]  ^{\left(  2\right)  }%
\end{array}
\right]  \cdot\left[
\begin{array}
[c]{c}%
A^-_{1}\\[5pt]
A^-_{2}%
\end{array}
\right] \, , \label{eq19}%
\end{equation}
where $A^-_{1,2}$  are unknown scalars.

Combining Eqs.~(\ref{meteq}), (\ref{eq18}), and (\ref{eq19}), we obtain
a matrix equation which may be rearranged as
\begin{equation}
[ \underline{\underline{M}}(\kappa)] \cdot\left[
\begin{array}
[c]{c}%
A^+_{1}\\[5pt]
A^+_{2}\\[5pt]
A^-_{1}\\[5pt]
A^-_{2}%
\end{array}
\right]  =\left[
\begin{array}
[c]{c}%
0\\[5pt]
0\\[5pt]
0\\[5pt]
0
\end{array}
\right]\,.  
\label{coeff}
\end{equation}
For a nontrivial solution, the $4\times4$ matrix $\left[
\underline{\underline{M}}(\kappa)\right]  $\ must be singular, so that%
\begin{equation}
\det\left[  \underline{\underline{M}}(\kappa)\right]  =0 \label{eq:SPPdisp}%
\end{equation}
is the dispersion equation for SPP-wave propagation.

\section{Numerical Results and Discussion}\label{nrd}

A Mathematica\texttrademark~program
was written and implemented to solve (\ref{eq:SPPdisp}) to obtain $\kappa$ for specific values of $\gamma^+$ and $\gamma^-$.
The dispersion equation~(\ref{eq:SPPdisp}) was solved using the Newton-Raphson technique~\cite{Jaluria} for three different values of the
twist between the morphologically significant planes on either side of the metal slab; we chose
\begin{itemize}
\item[(i)]  $\gamma^-=\gamma^+$,
\item[(ii)]  $\gamma^-=\gamma^++90^\circ$, and 
\item[(iii)]   $\gamma^-=\gamma^++45^\circ$,
\end{itemize}
while  $\gamma^+$ was kept as a variable.
For each choice,
the boundaries of the metal slab were taken to be at $L_\pm=\pm7.5$, $\pm12.5$, $\pm25$, or $\pm45$~nm. These  selections adequately
represent the results of our investigation. 

The free-space wavelength was fixed at $\lambdao=633$~nm for all calculations. The metal was taken to be  bulk aluminum
($\epsmet=-56+21i$). The skin depth of aluminum at the chosen wavelength is $\Delta_{met}=\left\{{\rm Im}\left[\ko\sqrt{\epsmet}\right]\right\}^{-1}=
13.24$~nm, a quantity of interest
in relation to the thickness of the metal slab.  
The SNTF was chosen
to be made of titanium oxide \cite{FL2010,HWH}, with 
\begin{equation}
\left.\begin{array}{l}
\epsa(z)=[1.0443+2.7394 v(z)-1.3697 v^2(z)]^2\\[5pt]
\epsb(z)=[1.6765+1.5649 v(z)-0.7825 v^2(z)]^2\\[5pt]
\epsc(z)=[1.3586+2.1109 v(z)-1.0554 v^2(z)]^2\\[5pt]
\chi(z)=\tan^{-1}[2.8818\tan\chiv(z)]
\end{array}\right\}\,,
\end{equation}
where  $v(z)=2\chiv(z)/\pi$.
The angles $\tilde\chi_v$ and $\delta_v$ were taken to be 45$^\circ$ and 30$^\circ$, respectively,
for all results presented here. 

\begin{figure}[!ht]
\begin{center}$
\begin{array}{cc}
\includegraphics[width=2.85in]{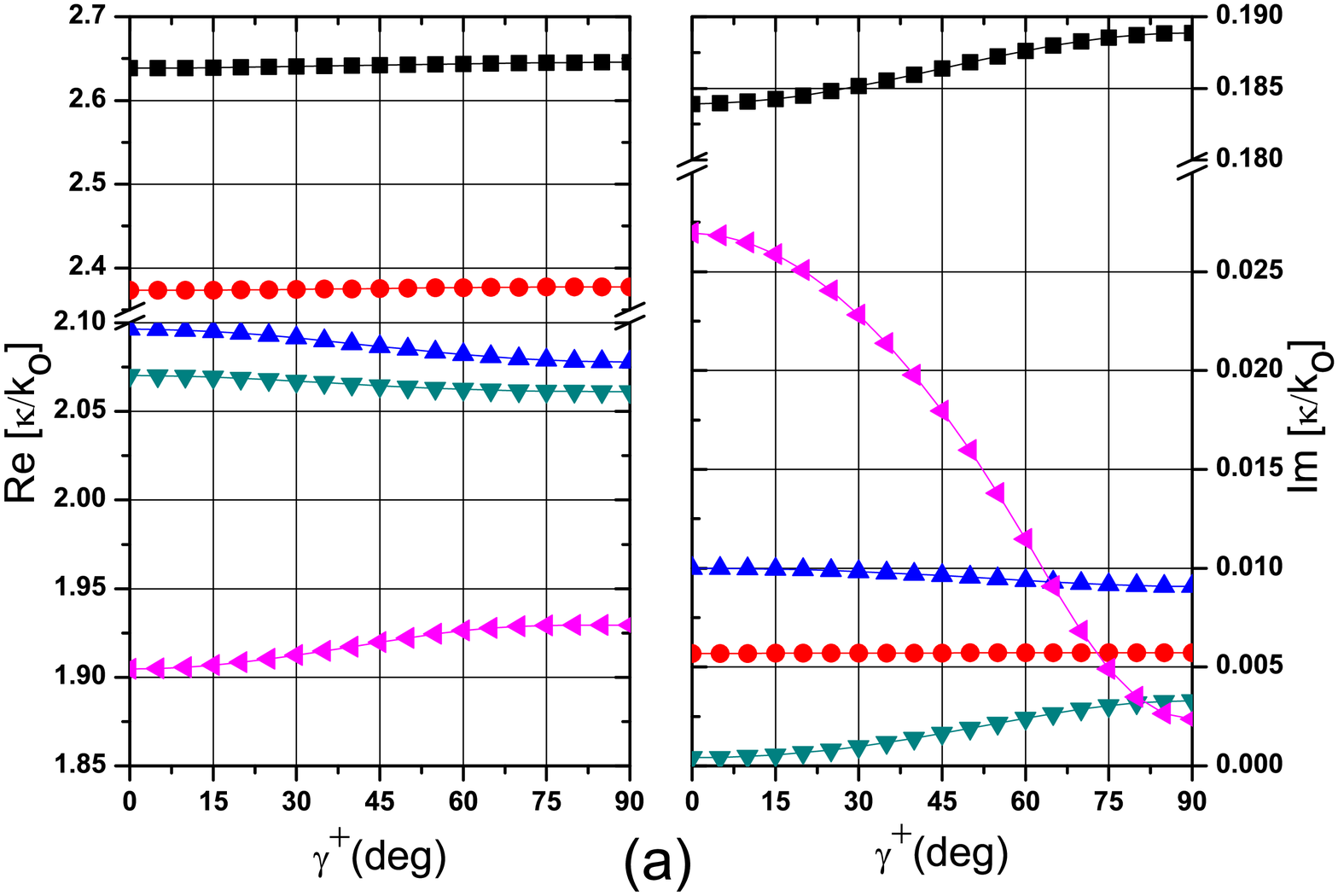}  \includegraphics[width=2.85in]{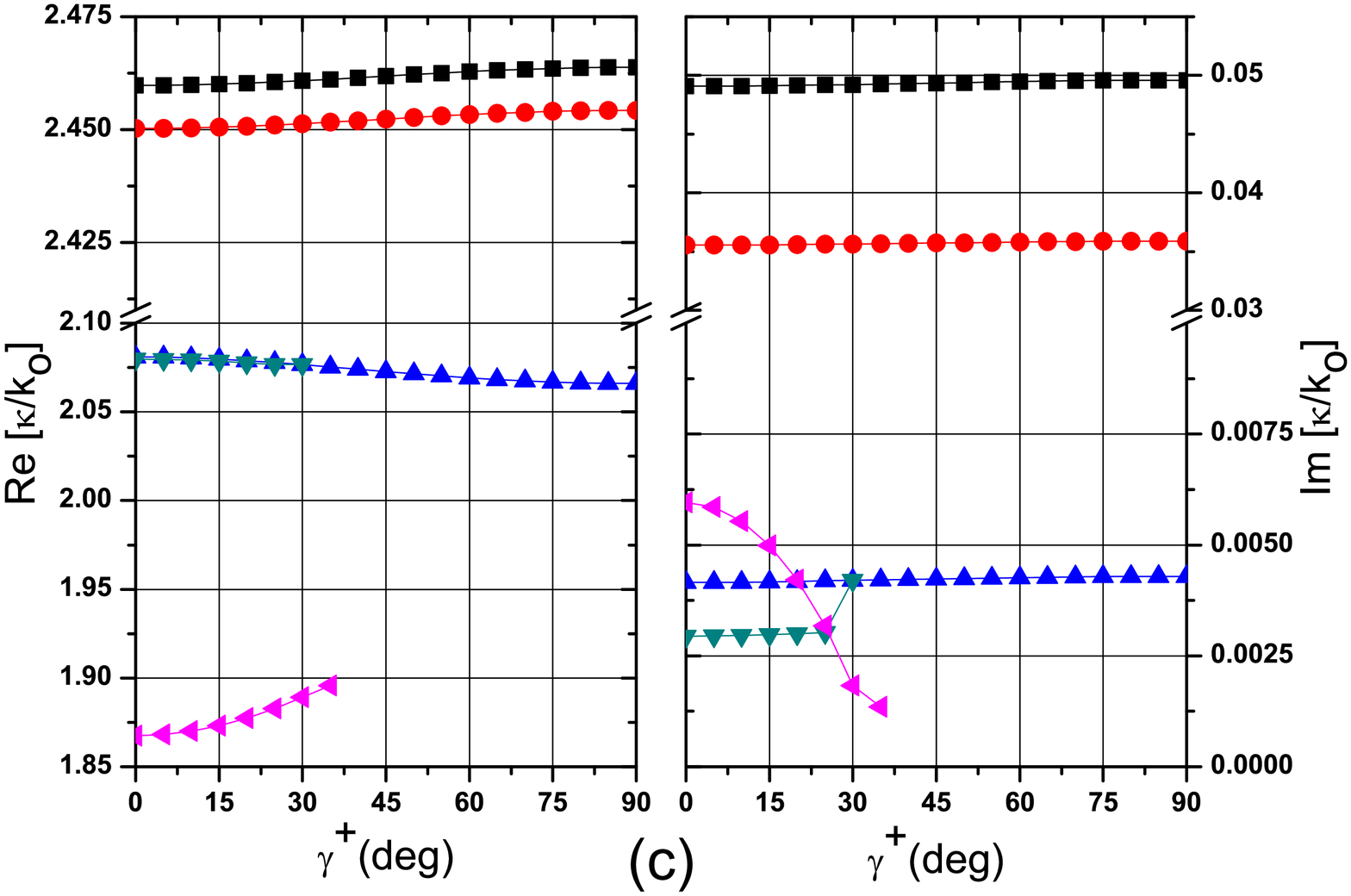}\\
\includegraphics[width=2.85in]{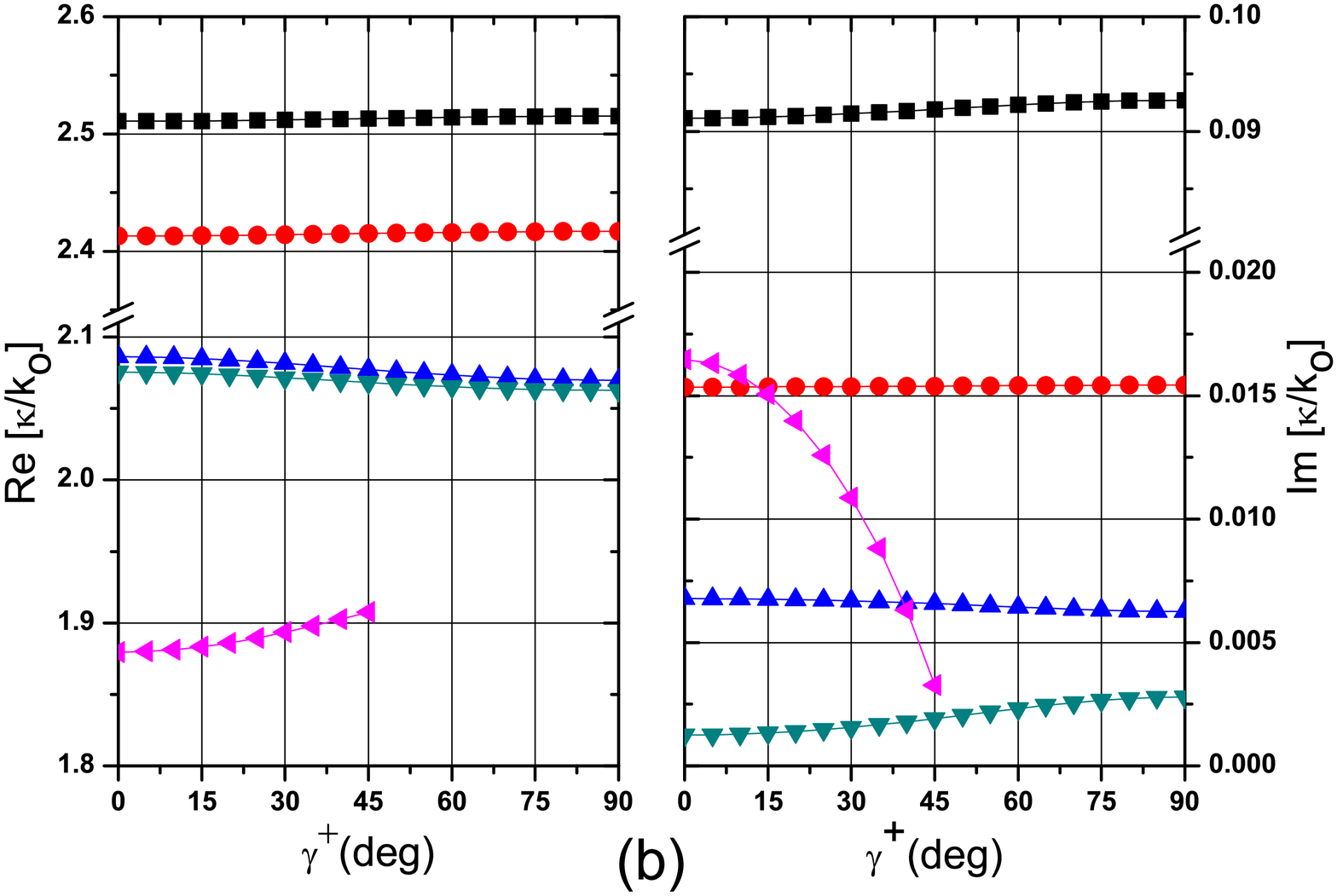} \includegraphics[width=2.85in]{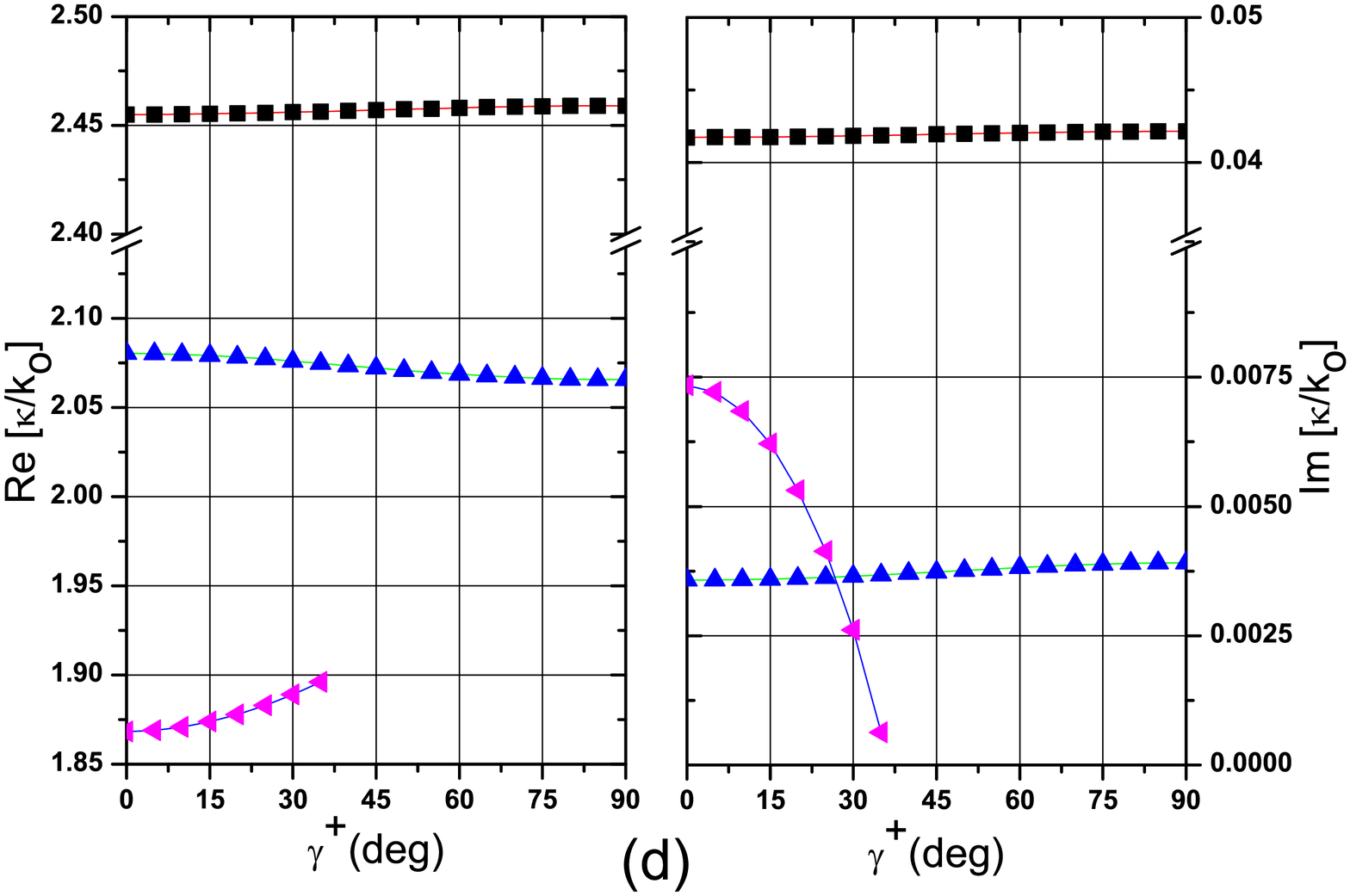}

\end{array}$
\caption{Variation of real and imaginary parts of $\kappa/{\ko}$ with $\gamma^+$, when $\gamma^-=\gamma^+$. (a) $L_\pm=\pm7.5$~nm, (b) $L_\pm=\pm12.5$~nm, (c) $L_\pm=\pm25$~nm, and (d) $L_\pm=\pm45$~nm. }
\label{kappa0}
\end{center}
\end{figure}

\begin{figure}[!ht]
\begin{center}$
\begin{array}{cc}
\includegraphics[width=2.25in]{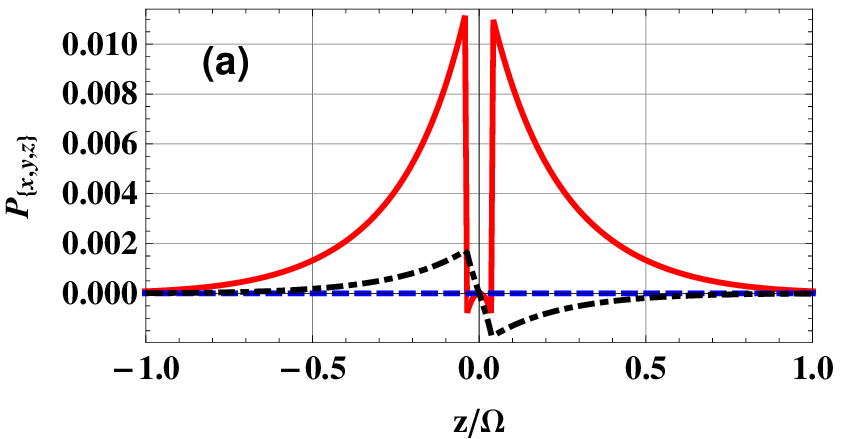} \includegraphics[width=2.25in]{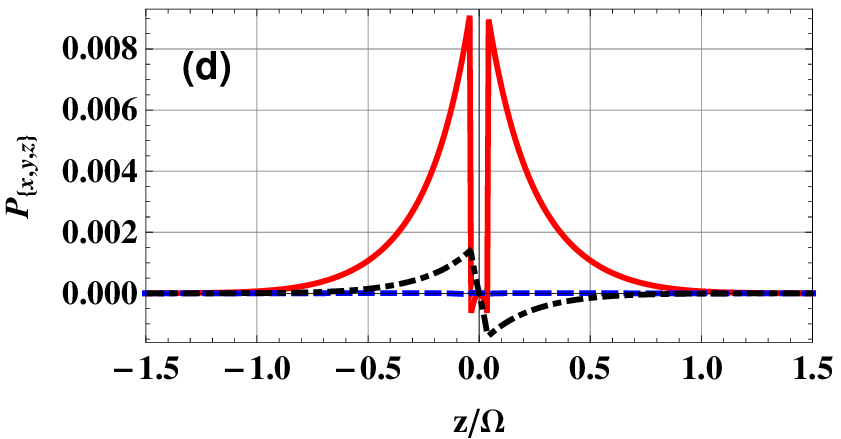}\\
\includegraphics[width=2.25in]{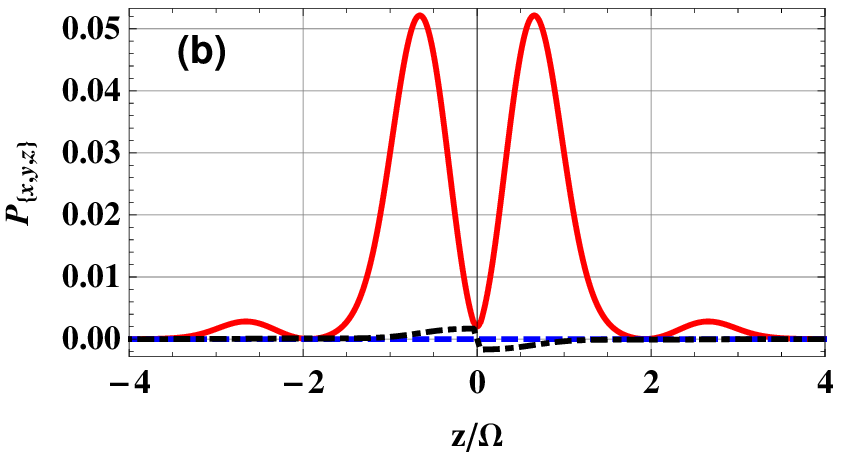} \includegraphics[width=2.25in]{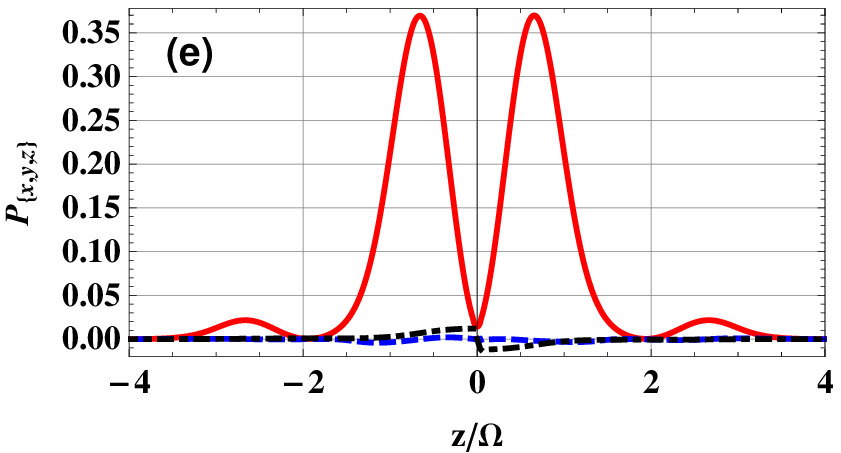}\\
\includegraphics[width=2.25in]{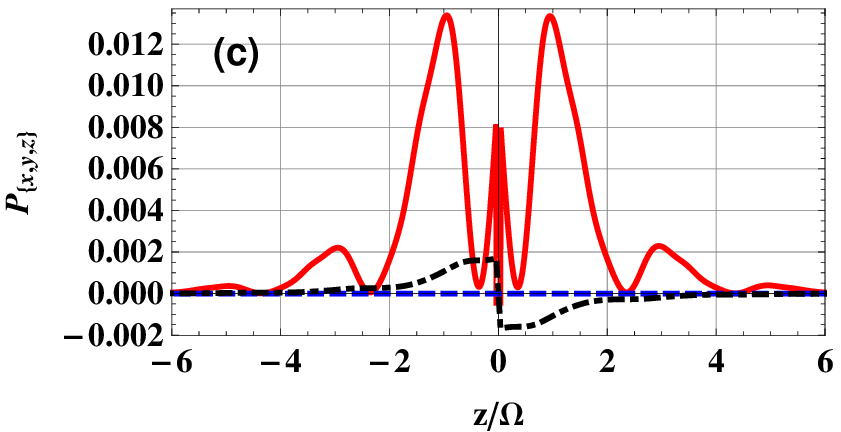} \includegraphics[width=2.25in]{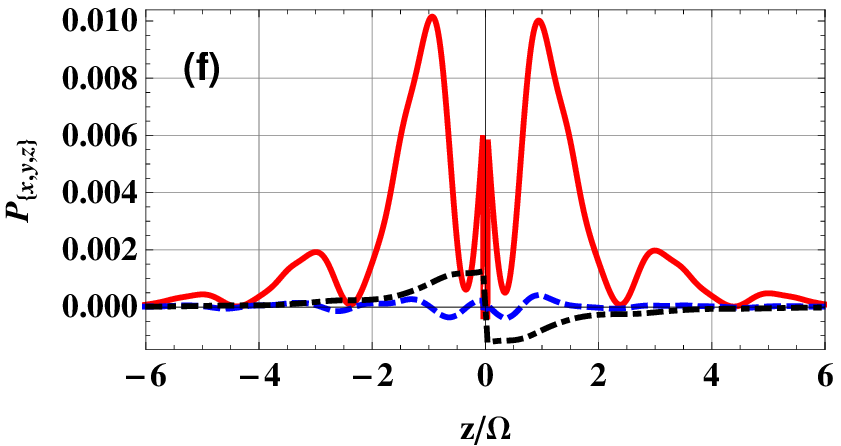}

\end{array}$
\caption{Variation of the Cartesian components of the time-averaged
Poynting vector ${\bf P}(x,z)$ along the $z$ axis when $x=0$, $L_\pm=\pm7.5~ \rm{nm}$, and $\gamma^-=\gamma^+$. (a-c) $\gamma^+=0$, and (d-f) $\gamma^+=25^\circ$. (a) $\kappa/{\ko}=2.6387+i 0.1839$, (b) $\kappa/{\ko}=2.0964+i 0.009997$, (c) $\kappa/{\ko}=1.9048+i0.02696$, (d) $\kappa/{\ko}=2.6399 + i0.1848$, (e) $\kappa/{\ko}=2.09285 +i0.00988$, and (f) $\kappa/{\ko}=1.9103+ i0.02405$. The $x$-, $y$- and $z$-directed components of ${\bf P}(x,z)$ are represented by solid red, dashed blue, and chain-dashed black lines, respectively. }
\label{field0L7.5}
\end{center}
\end{figure}

\begin{figure}[!ht]
\begin{center}$
\begin{array}{cc}
\includegraphics[width=2.25in]{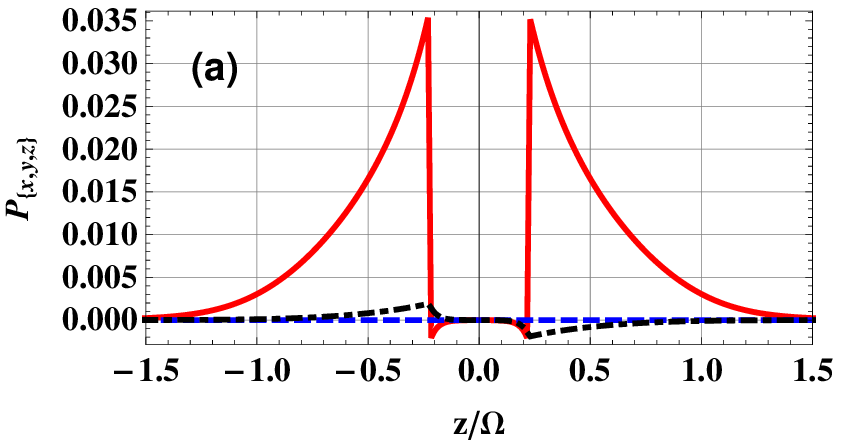} \includegraphics[width=2.25in]{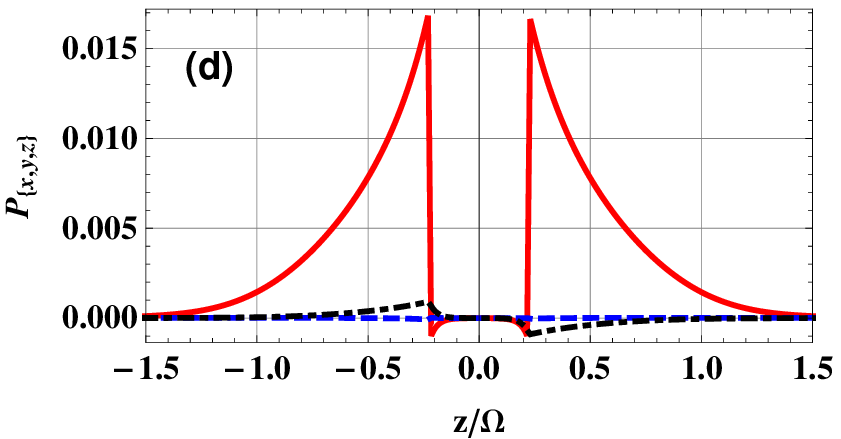}\\
\includegraphics[width=2.25in]{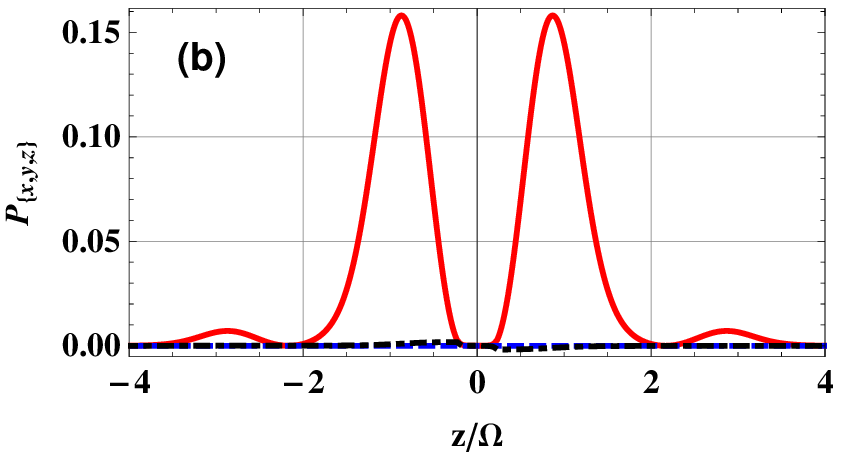} \includegraphics[width=2.25in]{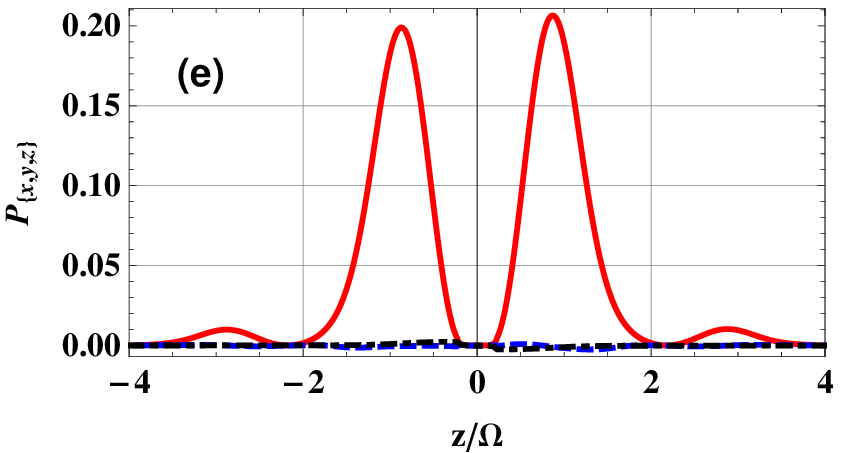}\\
\includegraphics[width=2.25in]{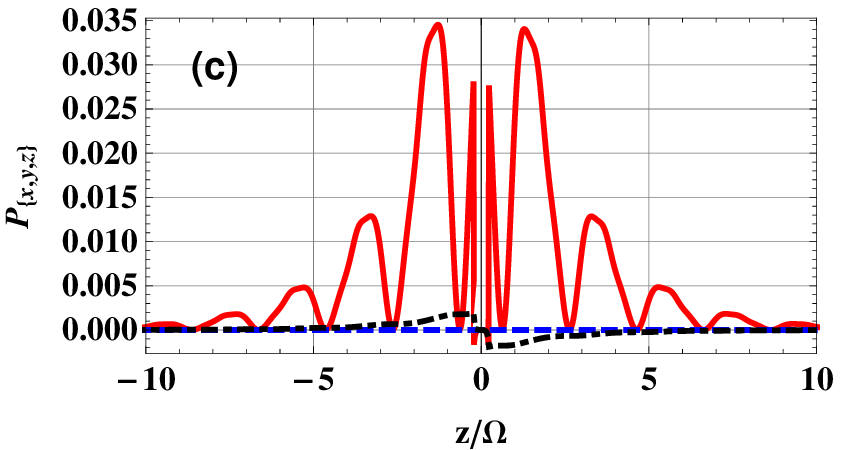} \includegraphics[width=2.25in]{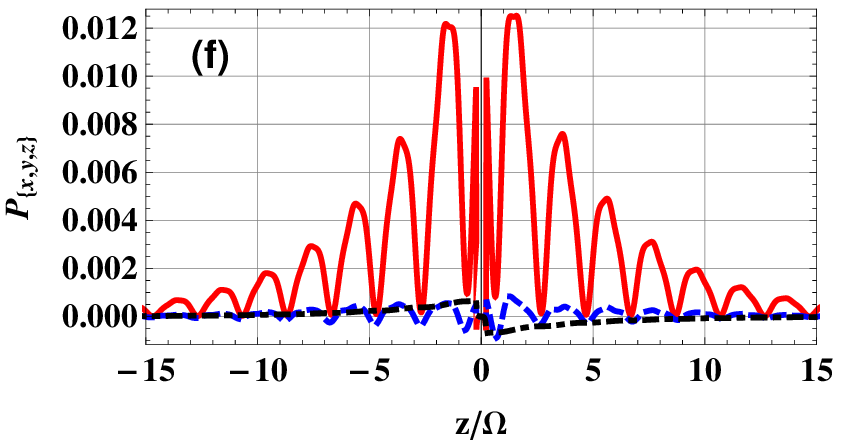}
\end{array}$
\caption{Same as Fig.~\ref{field0L7.5} except for $L_\pm=\pm45$~nm.  (a) $\kappa/{\ko}=2.4549+i 0.04173$, (b) $\kappa/{\ko}=2.08034+i 0.003574$, (c) $\kappa/{\ko}=1.8683 + i0.00734$, (d) $\kappa/{\ko}=2.4557+ i0.04181$, (e) $\kappa/{\ko}=2.0773 +i0.00363$, and (f) $\kappa/{\ko}=1.8830 + i0.00413$.  }
\label{field0L45}
\end{center}
\end{figure}

\subsection{$\gamma^-=\gamma^+$}\label{nrd0}

Let us begin with the case when the morphologically significant planes of the SNTF on both sides of the metal slab are aligned with each other
and make an angle $\gamma^-=\gamma^+$ with respect to the direction of SPP-wave propagation in the $xy$ plane. This case had
been briefly studied in a predecessor communication \cite{FL2010}, but is now discussed in detail.

The real and imaginary parts of $\kappa/{\ko}$ which satisfies Eq.~(\ref{eq:SPPdisp}) are presented in Fig.~\ref{kappa0} 
as functions of $\gamma^+\in\left[0^\circ,90^\circ\right]$; by virtue of symmetry, the solutions for $180^\circ+\gamma^+$ and
$360^\circ-\gamma^+$ are the same as for $\gamma^+$.
For the thinnest metal slab ($L_\pm=\pm7.5$~nm), the solutions are organized in five branches which span the entire range of $\gamma^+$. 
As the metal slab thickens to 25~nm ($L_\pm=\pm12.5$~nm), the number of branches does not change, but only four of those branches
span the entire range of $\gamma^+$ and one is confined to $\gamma^+\in \left[0^\circ, 49^\circ\right]$. Further thickening of
the metal slab to 50~nm
($L_\pm=\pm25$~nm) leads to five values of $\kappa$ satisfying the dispersion equation only in the range $\gamma^+\in \left[0^\circ, 35^\circ\right]$, four 
in the range $\gamma^+\in \left(35^\circ, 37^\circ\right]$, and three in the range $\gamma^+\in \left(37^\circ, 90^\circ\right]$. Finally,
for a 90-nm thick metal slab
($L_\pm=\pm45~\rm{nm}$) only three solutions exist for  $\gamma\in \left[0^\circ, 36^\circ\right]$ and two for $\gamma\in \left(36^\circ, 90^\circ\right]$, these
solutions being the same as for a single metal/SNTF interface \cite{PartIV}. We conclude that, as the thickness of the metal slab  is increased,
the coupling between two metal/SNTF interfaces $z=L_-$ and $z=L_+$ decreases; ultimately, the two
interfaces decouple from each other when the thickness $\lmet$ significantly exceeds
twice the skin depth $\Delta_{met}$ in the metal.

The solutions in Fig.~\ref{kappa0} can be categorized into three sets. The first set comprises those solutions for which ${\rm Re}[\kappa/{\ko}]$ lies between $2.3$ and $2.7$. This set has two branches when the metal slab is 15-nm thick, both branches spanning the entire
range of $\gamma^+$. As the thickness $\lmet$ increases,  the two branches come closer
to each other and eventually merge completely.  The second set comprises solutions for which ${\rm Re}[\kappa/\ko]$
lies between $2.05$ and $2.1$. It also has two branches when $\lmet = 15$~nm, both branches coalescing into one branch as
the thickness increases. Complete merger of the two branches of the first set occurs at a value of $\lmet$ smaller than that for the
two branches of the first set. Regardless of the value of $\lmet$, solutions in the second set can be found over the
entire range of $\gamma^+$. The third set consists of solutions lying on just one branch ($1.85 <{\rm Re}[\kappa/\ko] < 1.95$), but the maximum
value of $\gamma^+$ on this branch decreases rapidly from $90^\circ$ as $\lmet$ increases from $15$~nm.

The foregoing categorization is also meaningful as the spatial profiles of the fields are very similar for both solutions (for a specific value
of $\gamma^+$) in the first two sets. The same remark can be made for the spatial profiles of the time-averaged Poynting vector
\begin{equation}
{\bf P}(x,z)=\frac{1}{2}\,{\rm Re}\left[{\bf E}(x,z)\times {\bf H}^\ast(x,z)\right]\,.
\end{equation}
Representative plots of ${\bf P}(0,z)$ against $z$ are presented in
Figs.~\ref{field0L7.5} and \ref{field0L45} for combinations of (i) three values
of $\kappa$, one from each set, and (ii)  two values of $\gamma^+$. Whereas
 $\gamma^+=0$ was chosen because it represents the propagation of SPP waves in the morphologically significant plane of the SNTF
 on either side of the metal slab, $\gamma^+=25^\circ$ was chosen for the direction of SPP-wave propagation oblique to that plane.

Figure~\ref{field0L7.5} shows the spatial profiles of ${\bf P}(0,z)$ for $L_\pm=\pm7.5$~nm. The chosen values of
$\kappa/\ko$ are: (a)
 $2.6387+i 0.1839$, (b) $2.0964+i 0.009997$, and (c) $1.9048+i0.02696$ for $\gamma^+=0$; and (d) $2.6399 + i0.1848$, 
 (e) $2.09285 +i0.00988$,
 and (f) $1.9103+ i0.02405$ for $\gamma^+=25^\circ$. In order to determine these profiles, first $A^+_1$ was set equal to unity and then
 the remaining coefficients were found using Eq.~(\ref{coeff}) for all $\kappa/\ko$---except
for $\kappa/{\ko}=2.0964+i 0.009997$, for which $A^+_2=1$ was set   and the remaining coefficients were found using Eq.~(\ref{coeff}). 
The spatial profiles shown allow us to conclude that 
\begin{itemize}
\item[(i)] the SPP waves are bound strongly to 
both metal/SNTF interfaces, and
\item[(ii)] the power density mostly resides in the SNTF.
\end{itemize}

Representative calculated values of the penetration depth
$\Delta_{z}^\pm$ of the SPP wave into the metal, defined as the distance along the $z$ axis (in the metal slab) at which the amplitude of the
electric or magnetic field decays to $e^{-1}$ of its value at the nearest metal/SNTF interface $z=L_\pm$,  are tabulated in Table~1
for  $L_\pm=\pm 7.5~ \rm{nm}$. The penetration depths for all SPP waves are   $\sim12.7~\rm{nm}$, which confirms strong coupling
of the two metal/SNTF interfaces. This is not surprising since $\Delta_{z}^-=\Delta_{z}^+$ is very close to $\Delta_{met}$.

\begin{table}[width=\columnwidth,!ht]
\begin{center}
\caption{Penetration depths $\Delta_{z}^+=\Delta_{z}^-$ for  $L_\pm=\pm 7.5~ \rm{nm}$ and $\gamma^-=\gamma^+$. The
solutions are numbered in descending values of ${\rm Re}[\kappa/\ko]$.}
\vspace{0.3cm}
\begin{tabular}{|c||c|c|c|c|c|}\hline
\multicolumn{6}{|c|} {$\Delta_{z}^+=\Delta_{z}^-$~(nm)} \\ \hline
&  \multicolumn{5}{|c|} {Solution} \\ 
\cline{2-6}
  $\gamma^+=\gamma^-$   & 1st  & 2nd & 3rd & 4th & 5th \\\hline
 $0^\circ$ & 12.5458 & 12.6558 & 12.7780 & 12.7882 & 12.8564 \\\hline
 $15^\circ$ & 12.5456 & 12.6557 & 12.7786 & 12.7886 & 12.8555 \\\hline
 $30^\circ$ & 12.5450 & 12.6553 & 12.7801 & 12.7896 & 12.8532 \\\hline
 $45^\circ$ & 12.5443 & 12.6548 & 12.7821 & 12.7907 & 12.8500 \\\hline
 $60^\circ$ & 12.5436 & 12.6543 & 12.7839 & 12.7916 & 12.8470 \\\hline
 $75^\circ$ & 12.5431 & 12.6539 & 12.7852 & 12.7921 & 12.8454 \\\hline
 $90^\circ$ & 12.5430 & 12.6538 & 12.7857 & 12.7922 & 12.8451\\\hline
\end{tabular}\label{table1}
\end{center}
\end{table}

The spatial profiles of ${\bf P}(0,z)$ for $L_\pm=\pm45$~nm
are presented in Fig.~\ref{field0L45} for six selected values of
$\kappa/\ko$ identified in the figure caption. 
These profiles were obtained after $A^+_1$ was set equal to unity, except that $A^+_2=1$ was used for $\kappa/{\ko}=2.08034+i 0.003574$. 
The spatial profiles of ${\bf P}(0,z)$ for $L_\pm=\pm12.5$~nm and $\pm25$~nm have not been shown here because the spatial profiles for these cases are similar to those shown in Figs.~\ref{field0L7.5} and \ref{field0L45}. 
Representative values of the penetration depths $\Delta_{z}^\pm$ for $L_\pm=\pm45~\rm{nm}$ are given in Table~2. The penetration depths for this case are of the same order as for $L_\pm=\pm7.5~\rm{nm}$. As the thickness of the metal slab is $90~\rm{nm}$, it can be safely concluded that the SPP waves propagating on the two metal/SNTF interfaces are not coupled to each other but are propagating independently.

\begin{table}[width=\columnwidth,!ht]
\begin{center}
\caption{Penetration depths $\Delta_{z}^+=\Delta_{z}^-$ for  $L_\pm=\pm 45~ \rm{nm}$ and $\gamma^-=\gamma^+$. The
solutions are numbered in descending values of ${\rm Re}[\kappa/\ko]$.}
\vspace{0.3cm}
\begin{tabular}{|c||c|c|c|}\hline
\multicolumn{4}{|c|} {$\Delta_{z}^+=\Delta_{z}^-$~(nm)} \\ \hline
&  \multicolumn{3}{|c|} {Solution} \\ 
\cline{2-4}
  $\gamma^+=\gamma^-$   & 1st  & 2nd & 3rd  \\\hline
$0^\circ$ & 12.6206 & 12.7843 & 12.8691\\\hline
 $15^\circ$ & 12.6205 & 12.7848 & 12.8669\\\hline
 $30^\circ$ & 12.6201 & 12.7860  & 12.8608\\\hline
 $45^\circ$ & 12.6196 & 12.7877 &\\\hline
 $60^\circ$ & 12.6192 & 12.7891& \\\hline
 $75^\circ$ & 12.6188 & 12.7901& \\\hline
 $90^\circ$ & 12.6187 & 12.7904&\\\hline
\end{tabular}\label{table2}
\end{center}
\end{table}


\begin{figure}[!ht]
\begin{center}$
\begin{array}{cc}
\includegraphics[width=2.85in]{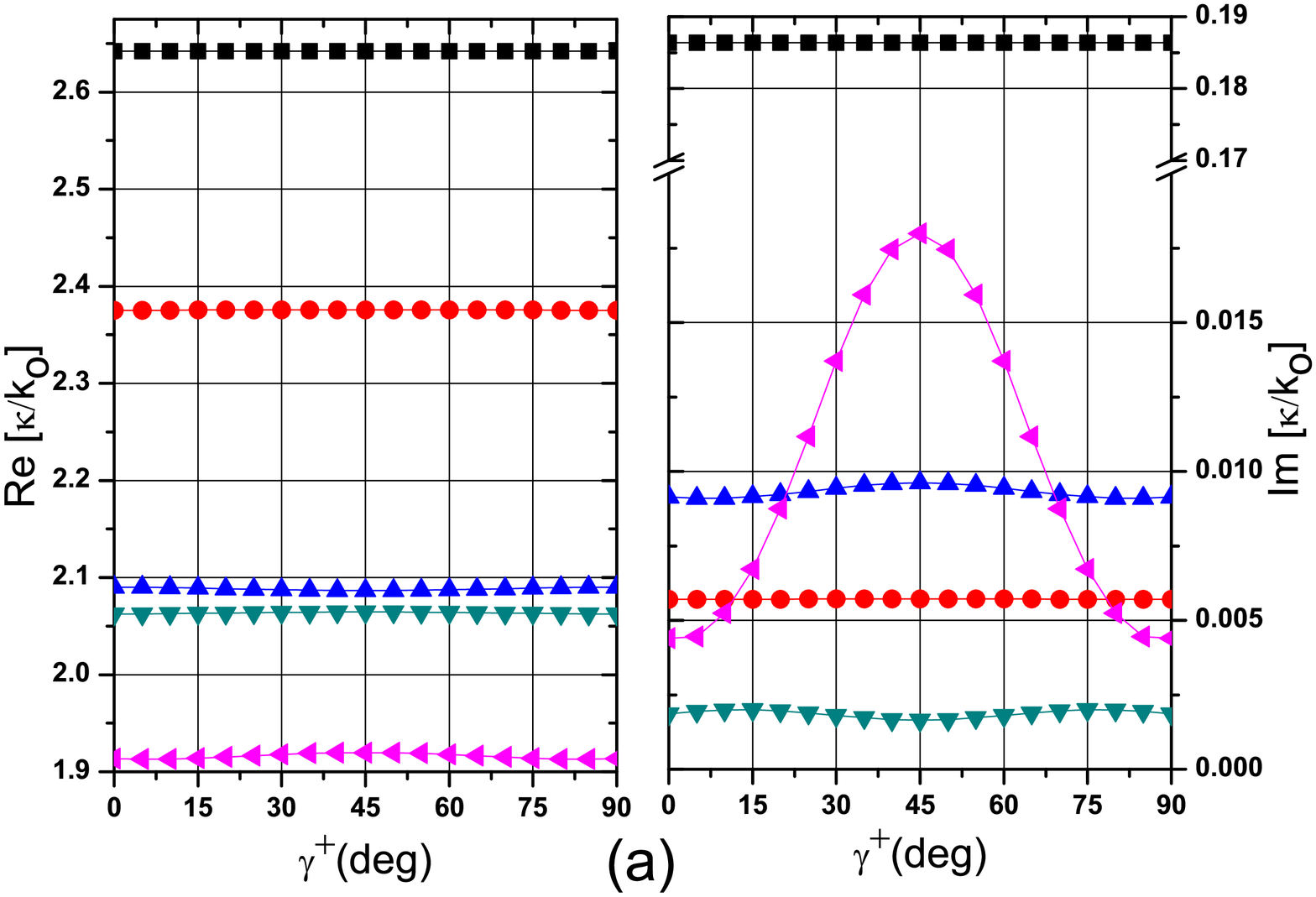}  \includegraphics[width=2.85in]{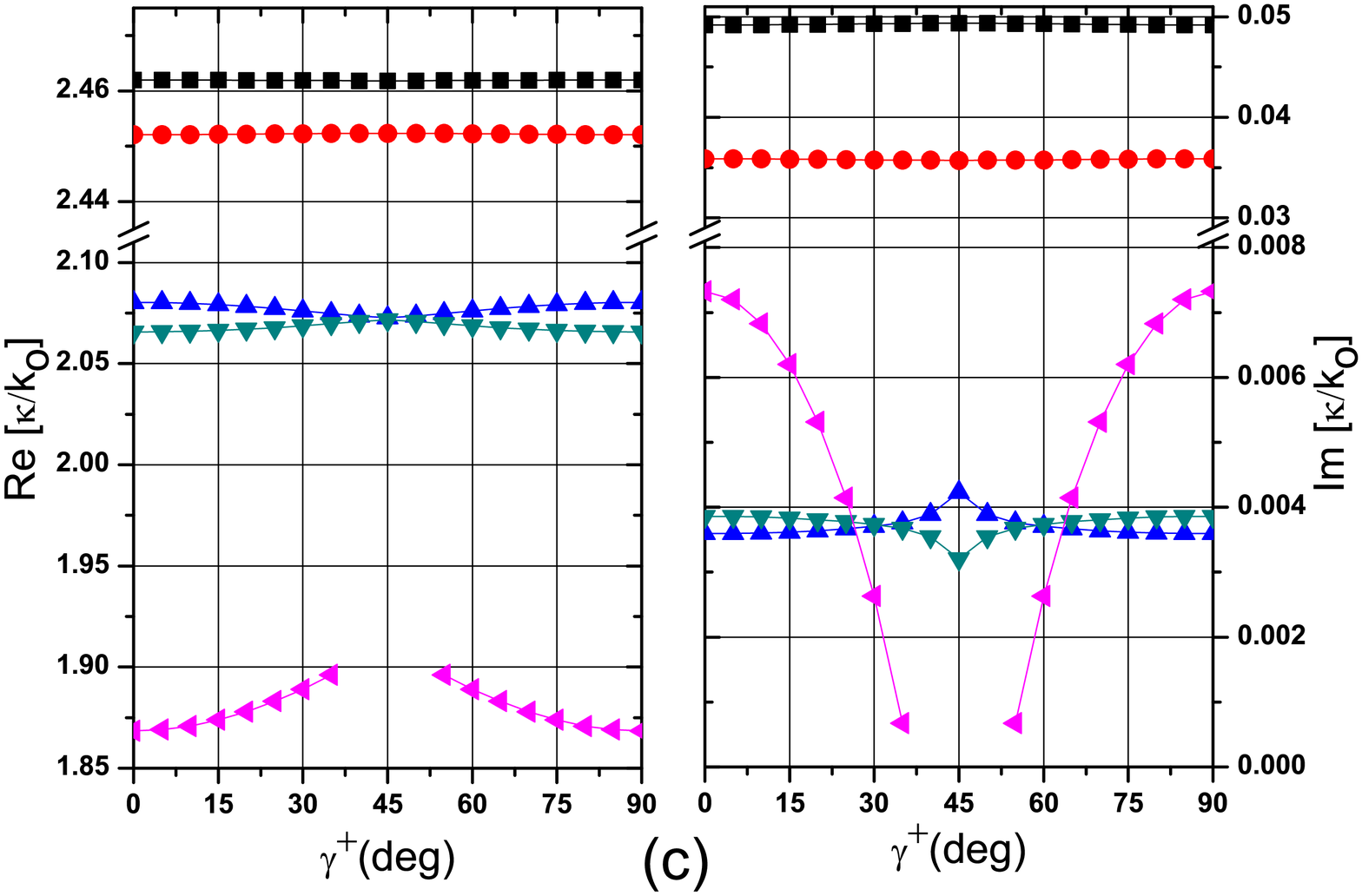}\\
\includegraphics[width=2.85in]{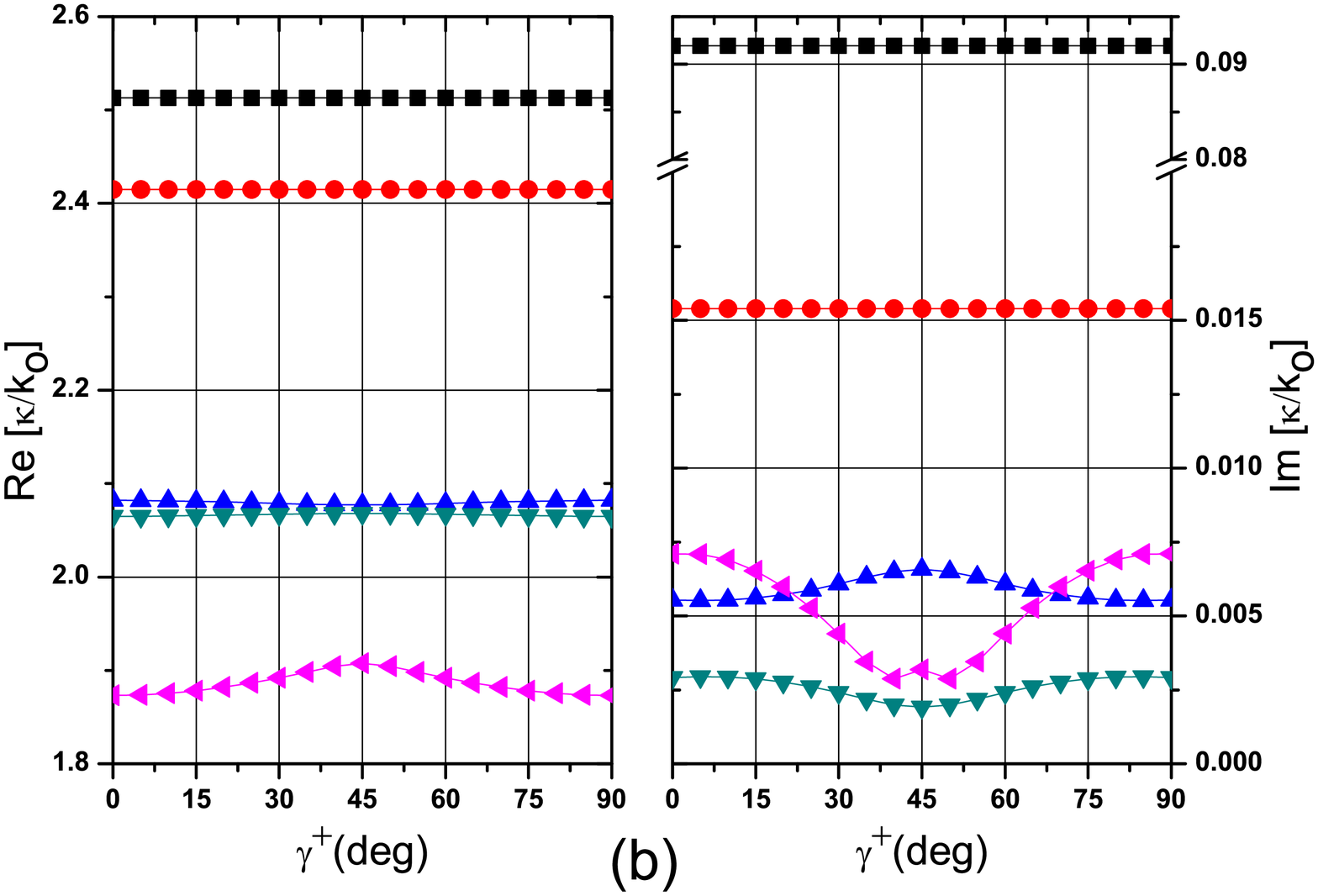} \includegraphics[width=2.85in]{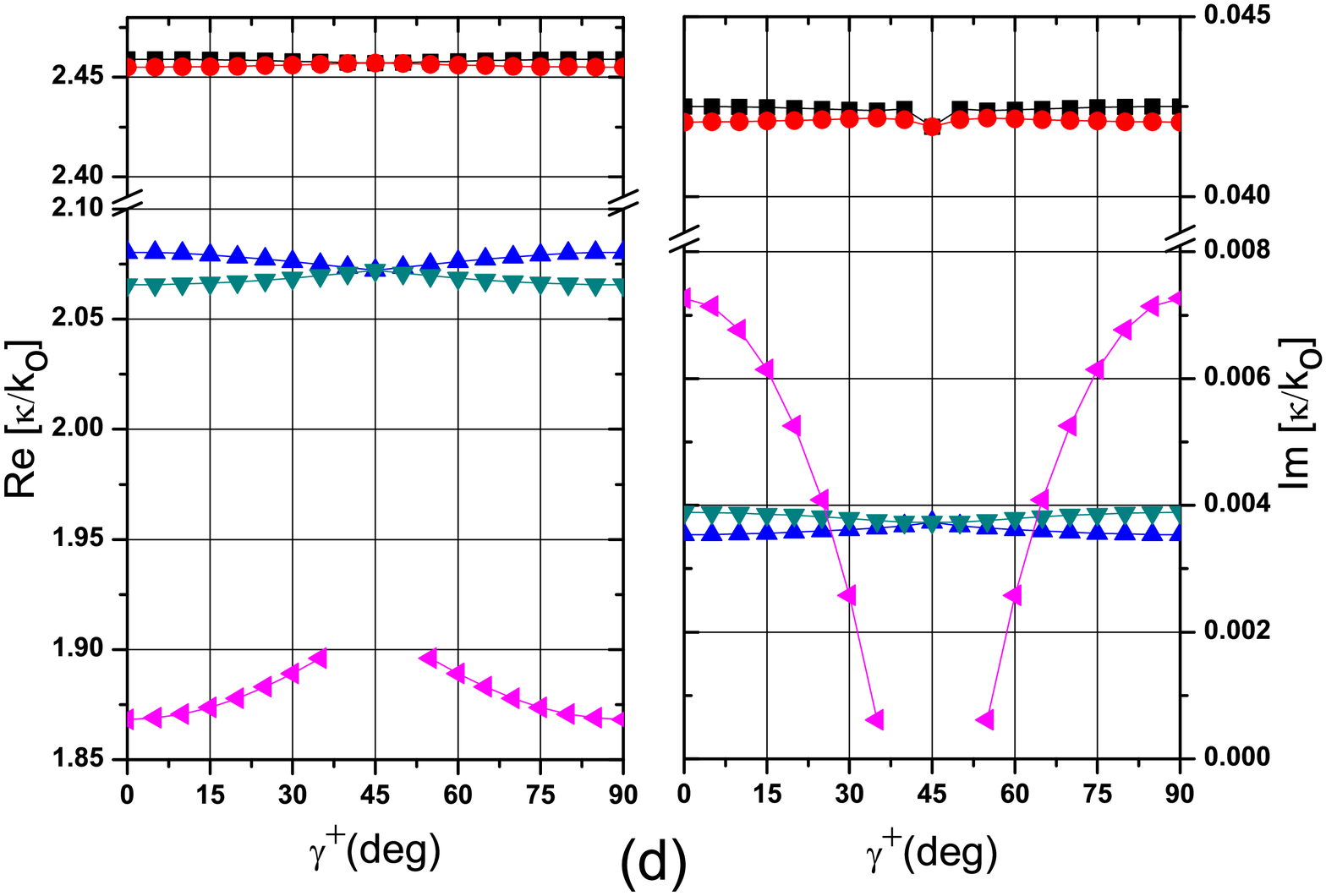}

\end{array}$
\caption{Same as Fig.~\ref{kappa0} except that $\gamma^-=\gamma^++90^\circ$. 
}
\label{kappa90}
\end{center}
\end{figure}

\begin{figure}[!ht]
\begin{center}$
\begin{array}{cc}
\includegraphics[width=2.25in]{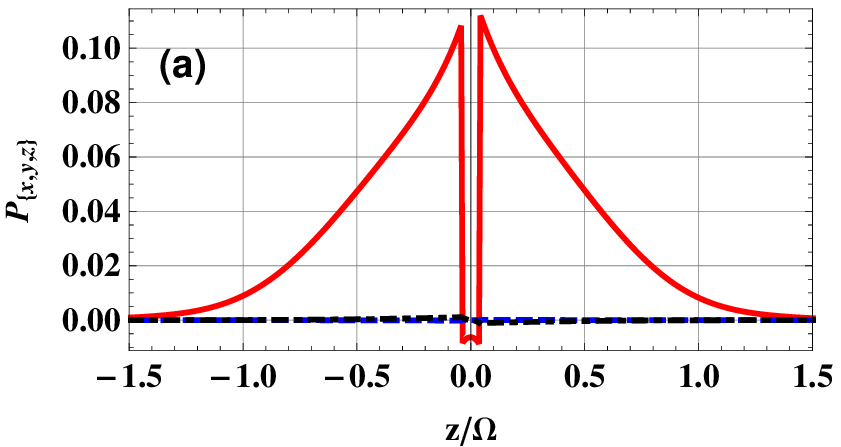} \includegraphics[width=2.25in]{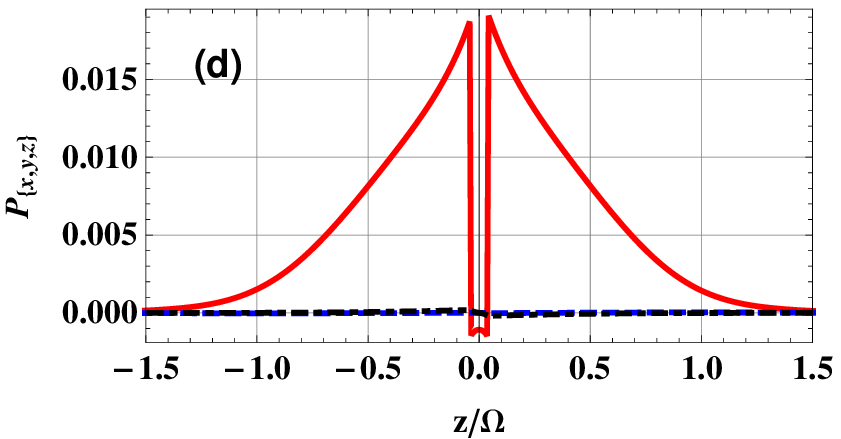}\\
\includegraphics[width=2.25in]{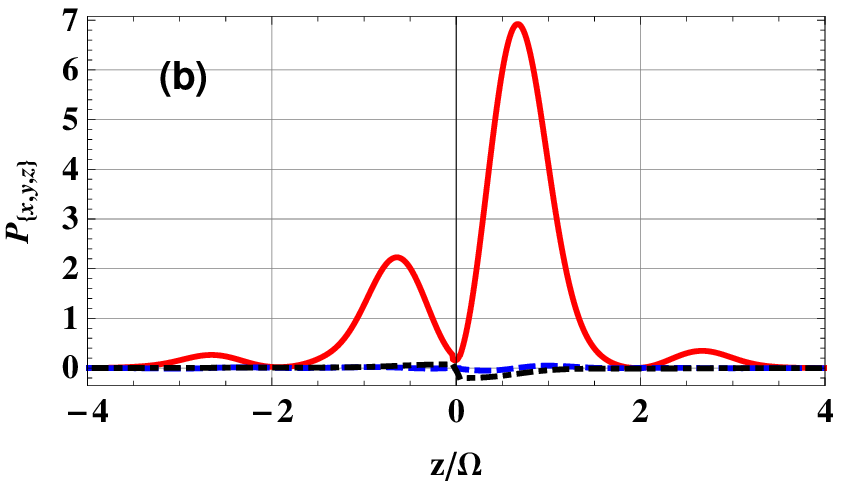} \includegraphics[width=2.25in]{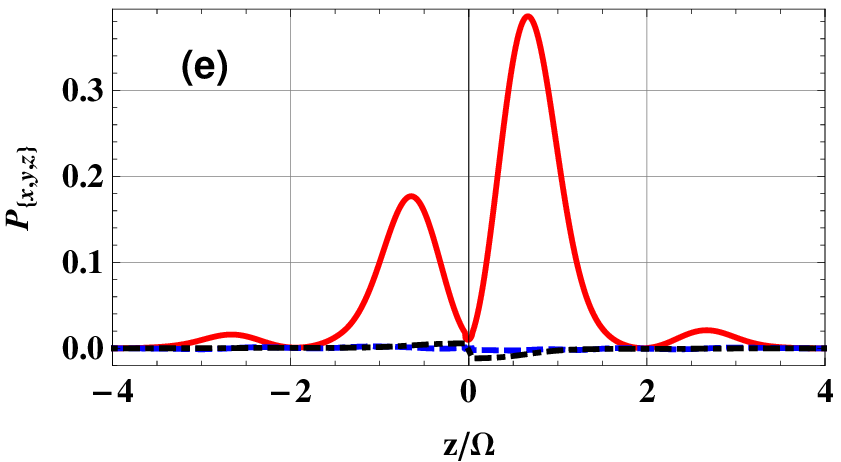}\\
\includegraphics[width=2.25in]{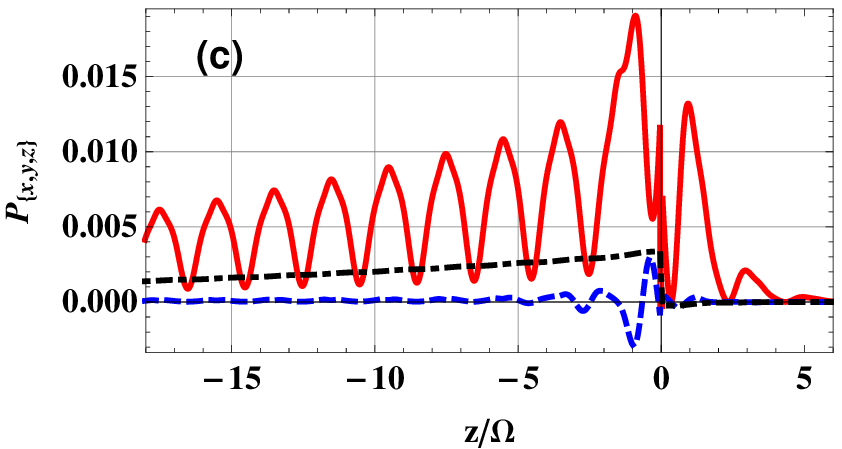} \includegraphics[width=2.25in]{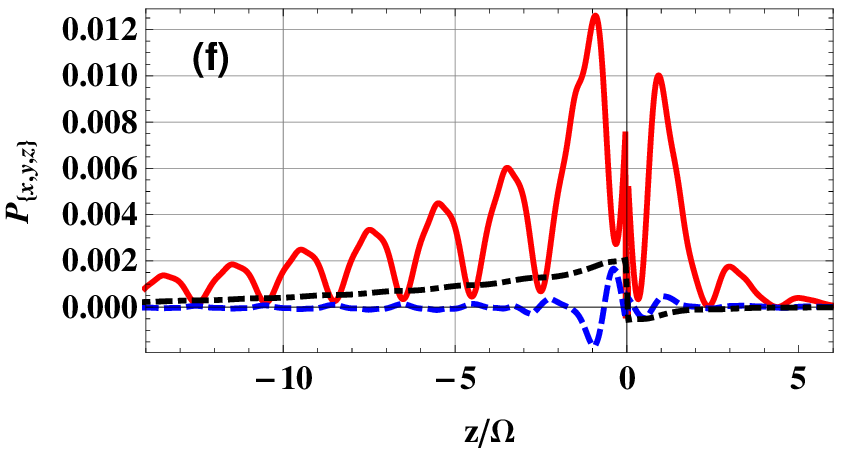}
\end{array}$
\caption{Variation of the Cartesian components of the time-averaged
Poynting vector ${\bf P}(x,z)$ along the $z$ axis when $x=0$, $L_\pm=\pm7.5~ \rm{nm}$, and $\gamma^-=\gamma^++90^\circ$. (a-c) $\gamma^+=0$, and (d-f) $\gamma^+=25^\circ$. The following values of $\kappa$ were chosen
for rough correspondence with those in Fig.~\ref{field0L7.5}:  (a) $\kappa/{\ko}=2.3753+i 0.005699$, (b) $\kappa/{\ko}=2.09013+i 0.009135$, (c) $\kappa/{\ko}=1.9133+i 0.004397$, (d) $\kappa/{\ko}=2.3756 + i0.005713$, (e) $\kappa/{\ko}=2.08797 +i0.009324$, and (f) $\kappa/{\ko}=1.9165 + i0.01117$. }
\label{field90L7.5}
\end{center}
\end{figure}

\begin{figure}[!ht]
\begin{center}$
\begin{array}{cc}
\includegraphics[width=2.25in]{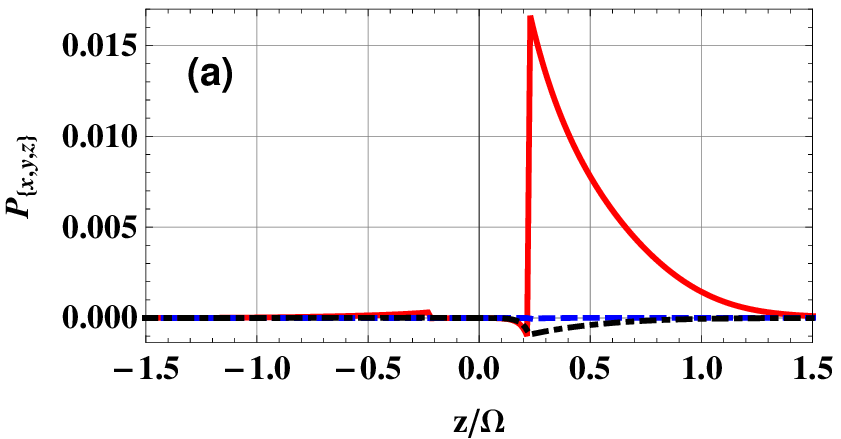} \includegraphics[width=2.25in]{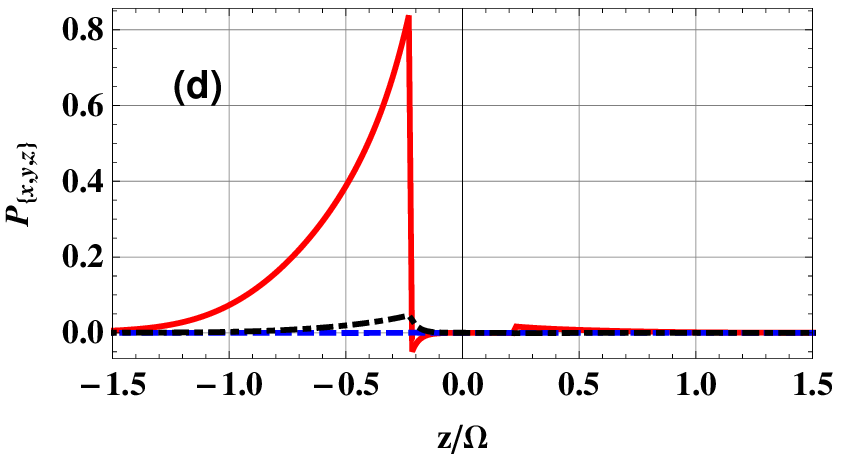}\\
\includegraphics[width=2.25in]{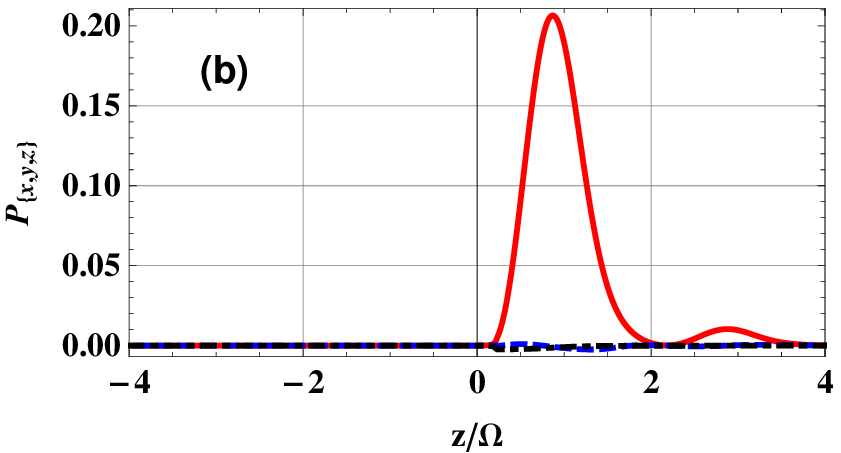} \includegraphics[width=2.25in]{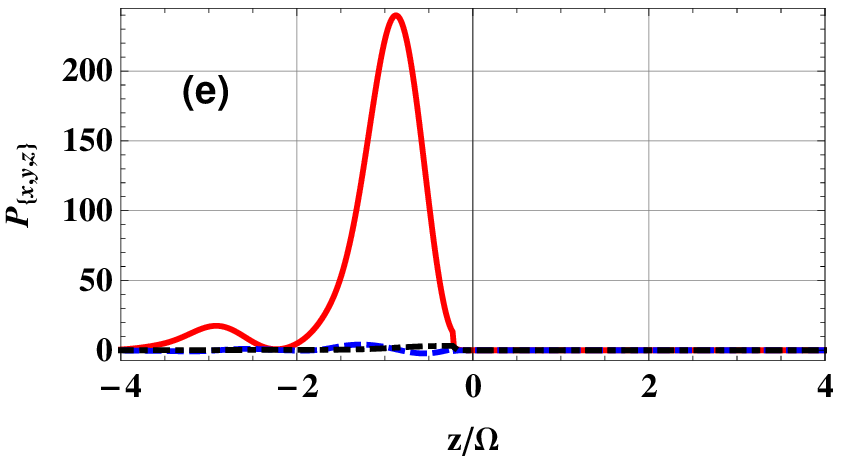}\\
\includegraphics[width=2.25in]{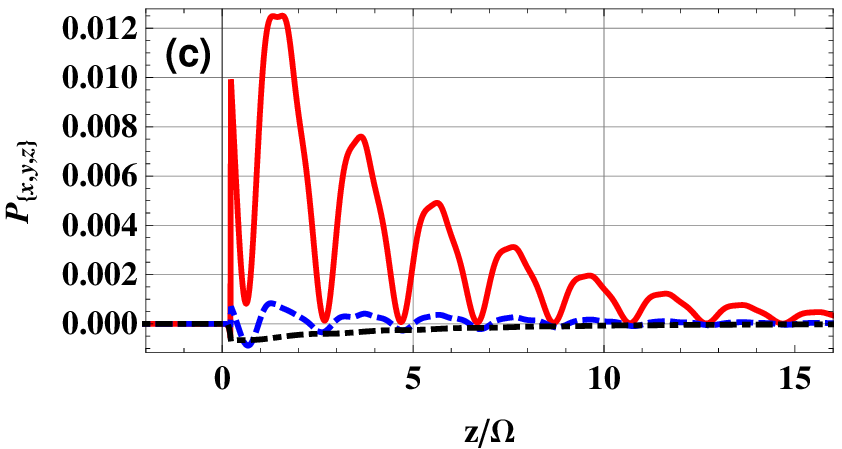} \includegraphics[width=2.25in]{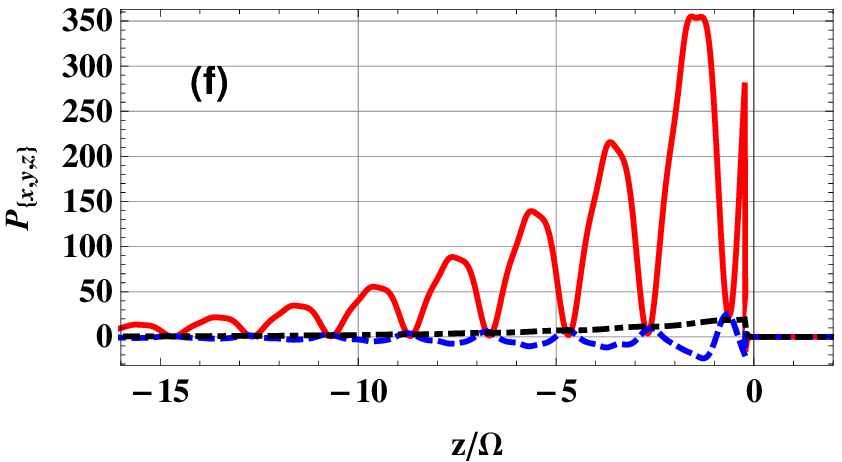}
\end{array}$
\caption{Variation of the Cartesian components of the time-averaged
Poynting vector ${\bf P}(x,z)$ along the $z$ axis when $x=0$, $L_\pm=\pm45~ \rm{nm}$, and $\gamma^-=\gamma^++90^\circ$.  
The following values of $\kappa$ and $\gamma^+$ were chosen to highlight the uncoupling of the two metal/SNTF interfaces, when the metal slab is sufficiently thick.
(a-e) $\gamma^+=25^\circ$ and (f) $\gamma^+=65^\circ$. (a) $\kappa/{\ko}=2.4558 + i0.04214$, (b) $\kappa/{\ko}=2.07725 + i0.003595$, (c) $\kappa/{\ko}=1.8830+ i0.004085$, (d) $\kappa/{\ko}=2.45833 + i0.0424391$, (e) $\kappa/{\ko}=2.06775 + i0.00381762$, and (f) $\kappa/{\ko}=1.8830+ i0.004085$. }
\label{field90L45}
\end{center}
\end{figure}

\subsection{$\gamma^-=\gamma^++90^\circ$}\label{nrd90}

The real and imaginary parts of $\kappa/k_\circ$ that satisfies the dispersion equation (\ref{eq:SPPdisp}) for $\gamma^-=\gamma^++90^\circ$ are given in Fig.~\ref{kappa90} as functions of $\gamma^+\in\left[0^\circ,90^\circ\right]$. Due to the symmetry of the problem, the solutions for $90^\circ\pm\gamma^+$, $180^\circ\pm\gamma^+$, $270^\circ\pm\gamma^+$ and $360^\circ-\gamma^+$ are the same as for $\gamma^+$. We found five solutions which span the whole range of $\gamma^+$ for $L_\pm=\pm7.5~\rm{and}~\pm12.5~\rm{nm}$.  However, when $L_\pm=\pm25~\rm{and}~\pm45~\rm{nm}$,
only four solutions exist for $\gamma\in \left(37^\circ, 53^\circ\right)$, but five for other values of $\gamma^+\in\left[0^\circ,37^\circ\right]\cup \left[53^\circ,90^\circ\right]$. 

For the thickest slab considered, we see from
comparing the results presented in Fig.~\ref{kappa90}(d) with those in Fig.~\ref{kappa0}(d)---which are same as that for a single metal/SNTF interface---that
the two metal/SNTF interfaces are actually uncoupled from each other.
The solutions in Fig.~\ref{kappa90}(d) represent SPP-wave propagation guided by a metal/SNTF interface with the direction of propagation in the $xy$ plane either making an angle $\gamma^+$ or $\gamma^-$with the morphologically significant plane of the SNTF.
 
As in Sec.~\ref{nrd0}, the solutions of the dispersion equation can be categorized into three sets with the same criterions as given in Sec.~\ref{nrd0}. However, in this case, the two branches in either of the first two sets do not merge into one branch as $L_{met}$ is increased. Instead, the two branches remain distinct, each holding for SPP-wave propagation guided by one of the two metal/SNTF interfaces uncoupled from the other metal/SNTF interface.
The single branch in the third set
 actually vanishes for mid-range values of $\gamma^+$, with the two parts of that branch signifying independent propagation guided by the two metal/SNTF interfaces. 
 
 Examination of the field profiles confirms the conclusions made in Sec.~\ref{nrd0} regarding the effect of the thickness of the metal slab. The asymmetry
 in the alignment of the morphologically significant planes for $z<L_-$ and $z>L_+$ is reflected in the spatial profiles of the time-averaged Poynting vector
 presented in Figs.~\ref{field90L7.5} and \ref{field90L45} for $\lmet=15$ and $90$~nm, respectively. Fig.~\ref{field90L7.5} also suggests that the coupling of the SPP-wave is directly proportional to the real part of wavenumber. So slower SPP waves are more strongly coupled to both interfaces. This observation is also confirmed by the spatial profiles of SPP waves when $\lmet=25$ and $50$~nm (not shown). Fig.~\ref{field90L45} shows the uncoupling of the two metal/SNTF interfaces when the metal slab is sufficiently thick.

\begin{figure}[!ht]
\begin{center}$
\begin{array}{cc}
\includegraphics[width=2.85in]{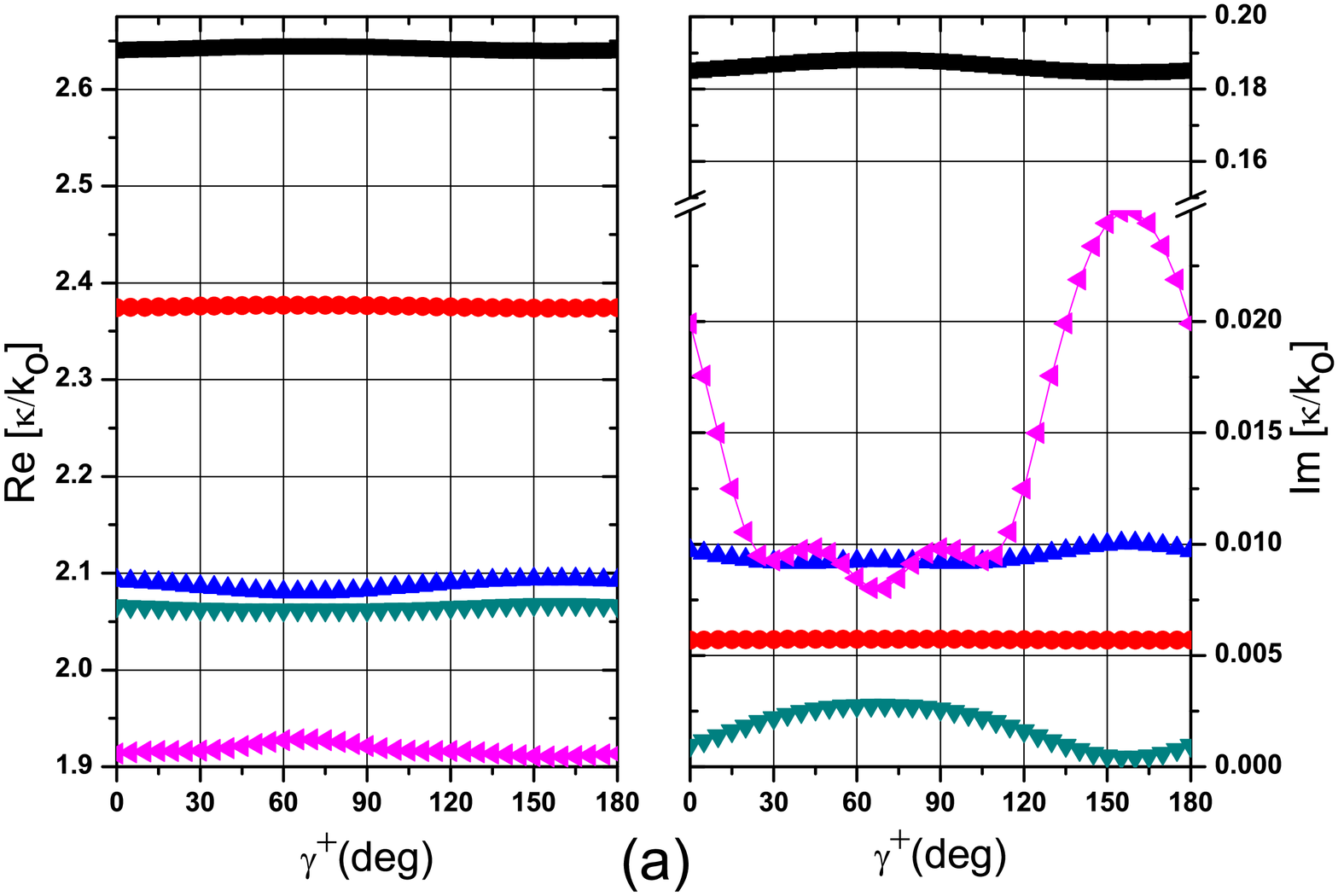}  \includegraphics[width=2.85in]{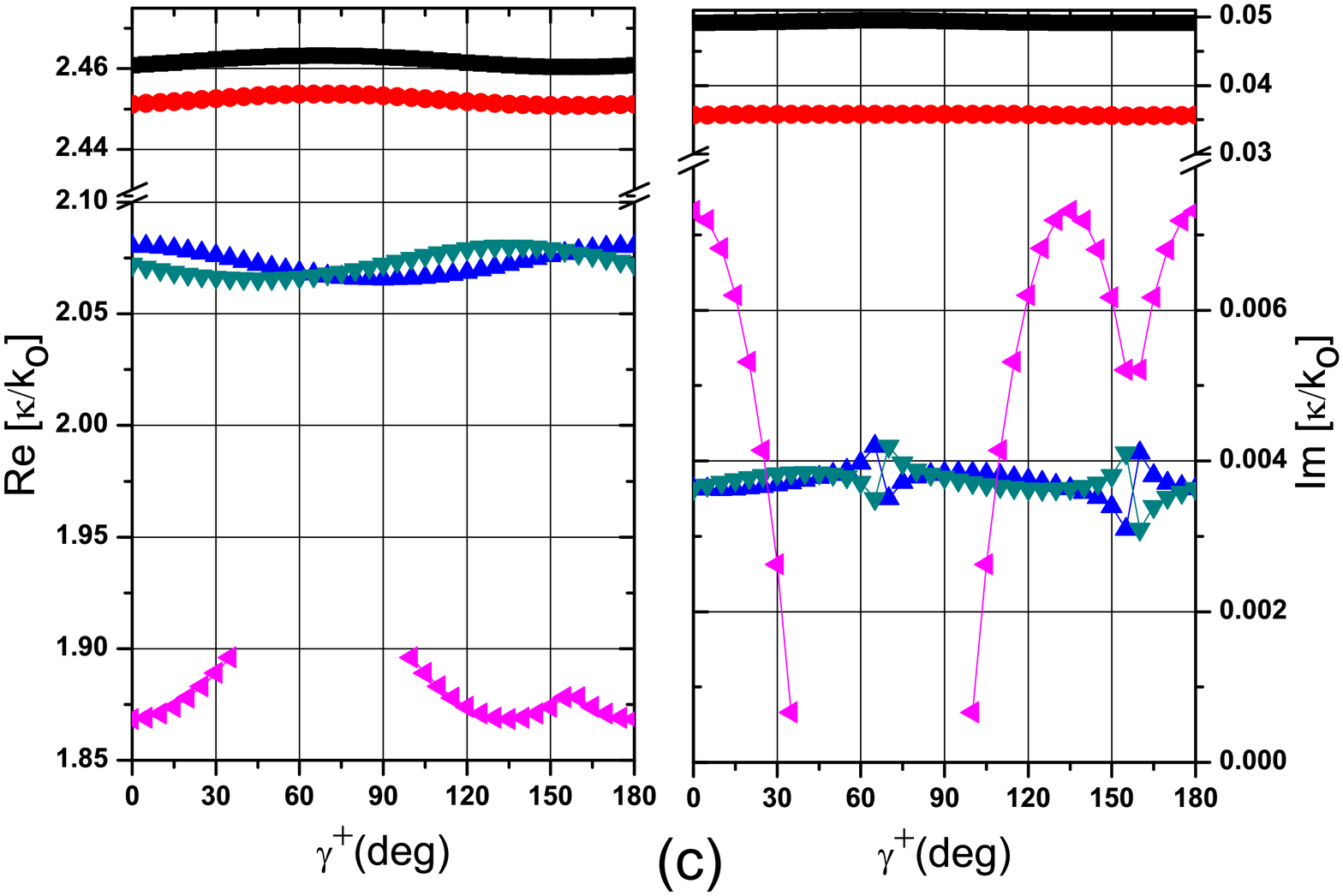}\\
\includegraphics[width=2.85in]{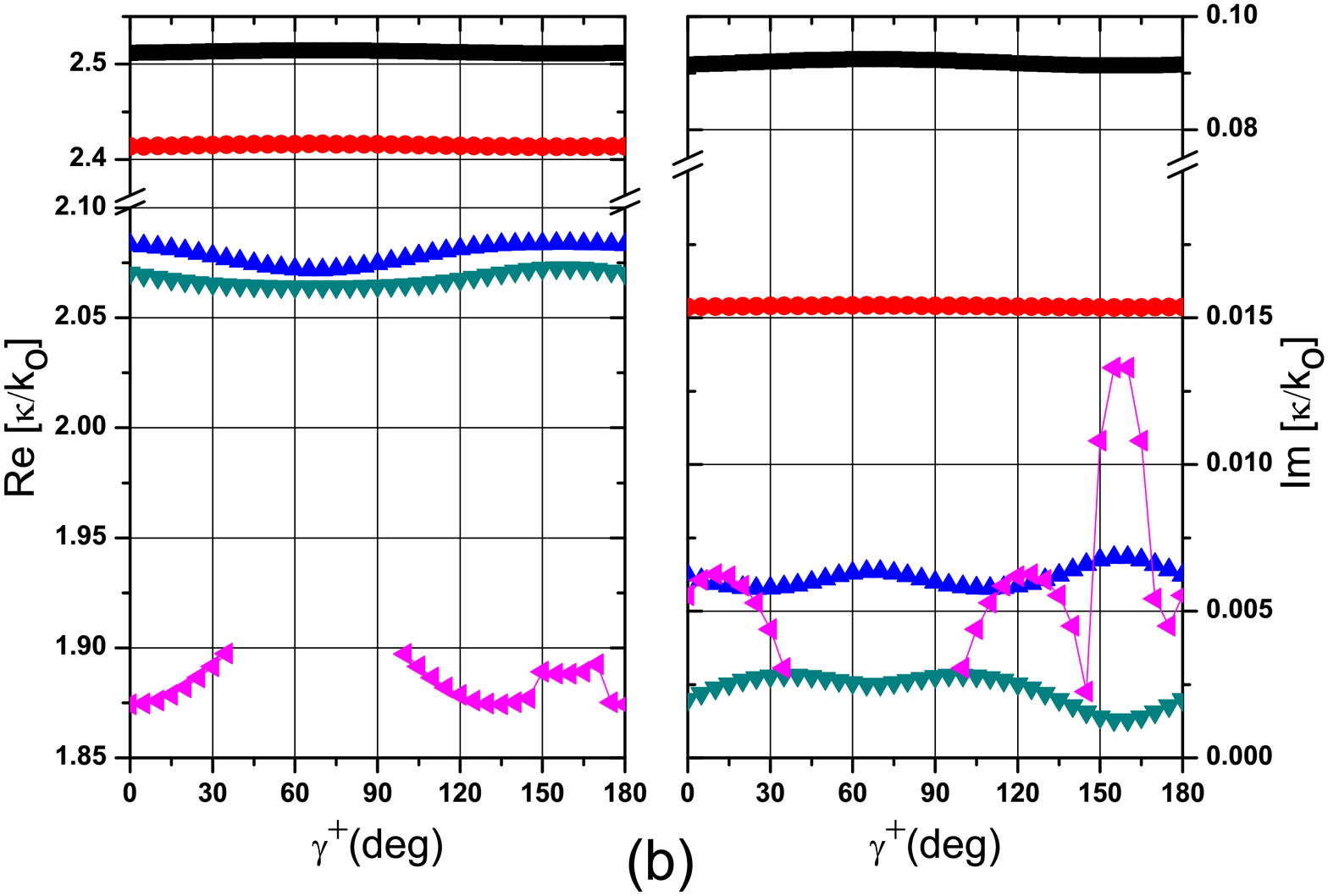} \includegraphics[width=2.85in]{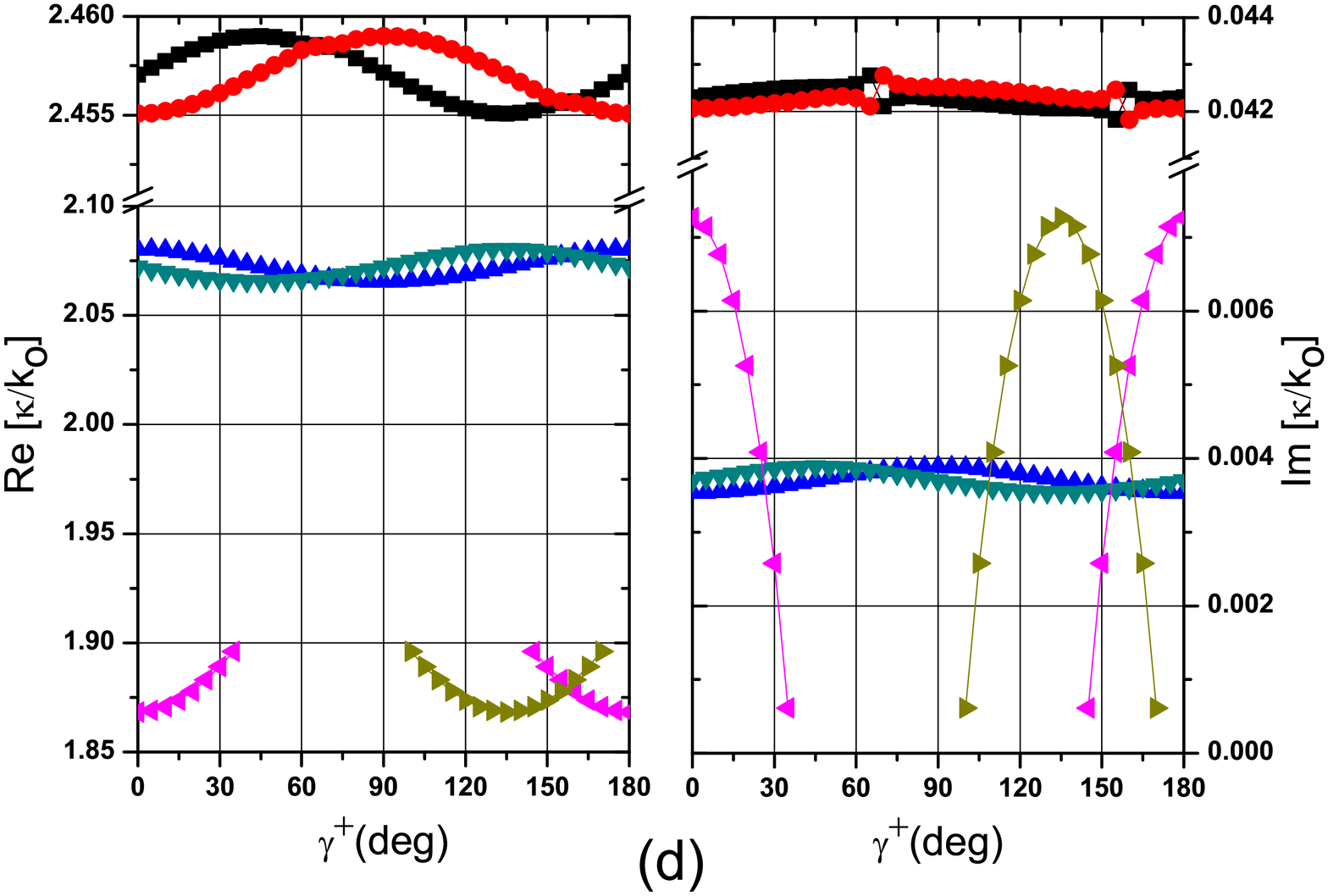}

\end{array}$
\caption{Same as Fig.~\ref{kappa0} except $\gamma^-=\gamma^++45^\circ$. }
\label{kappa45}
\end{center}
\end{figure}
 
\begin{figure}[!ht]
\begin{center}$
\begin{array}{ccc}
\includegraphics[width=2.25in]{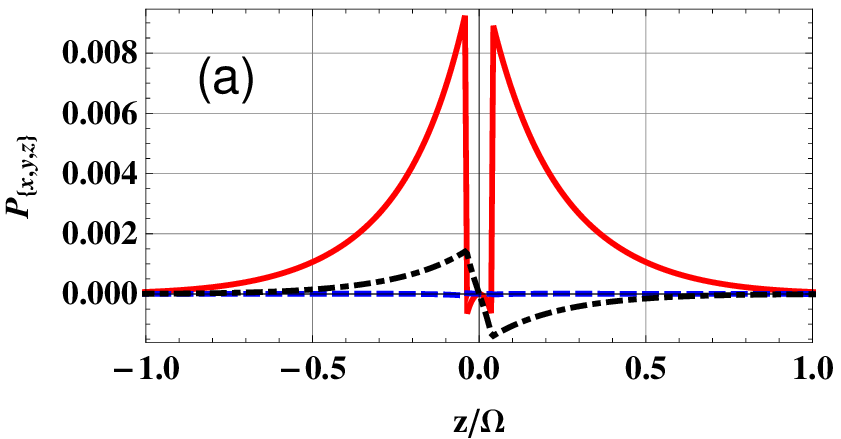}  \includegraphics[width=2.25in]{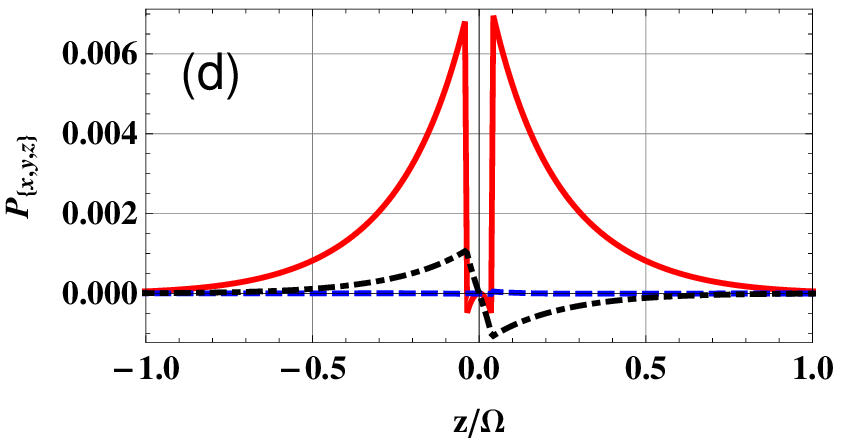} \\
\includegraphics[width=2.25in]{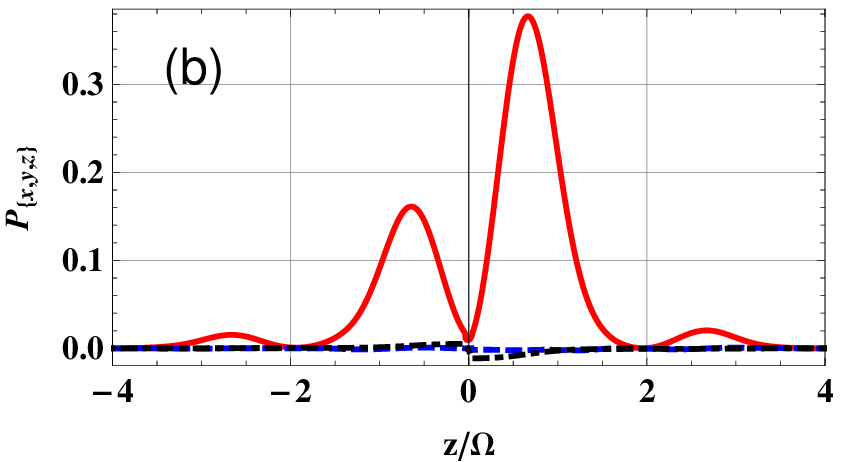} \includegraphics[width=2.25in]{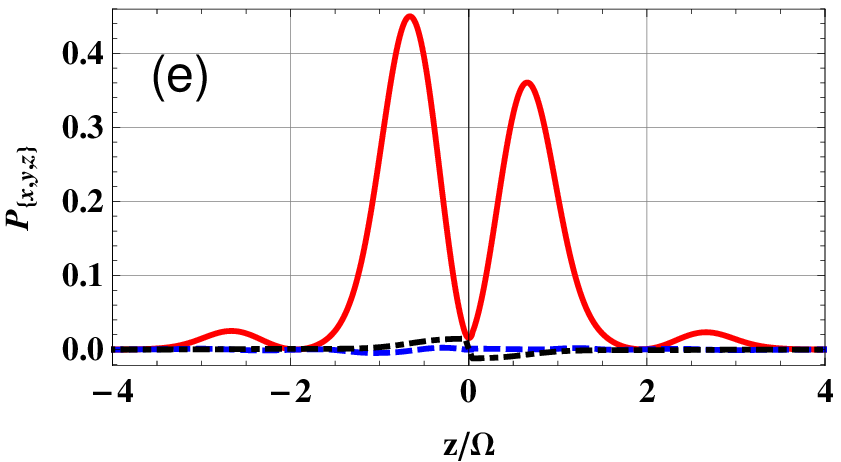} \\
\includegraphics[width=2.25in]{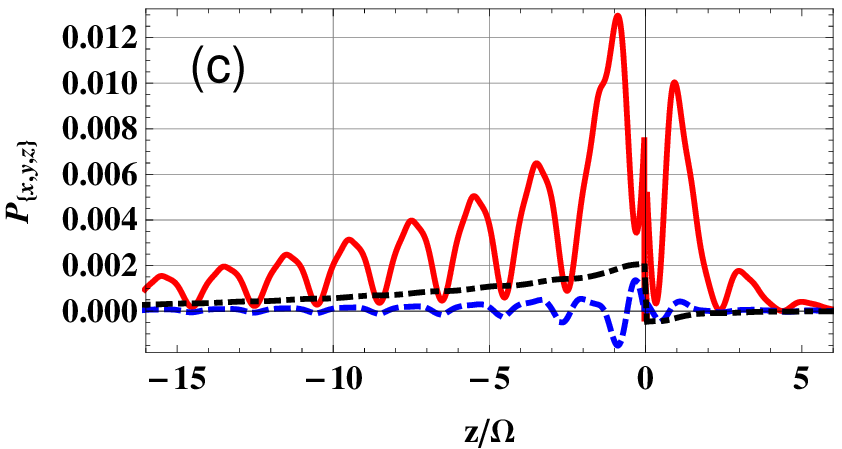} \includegraphics[width=2.25in]{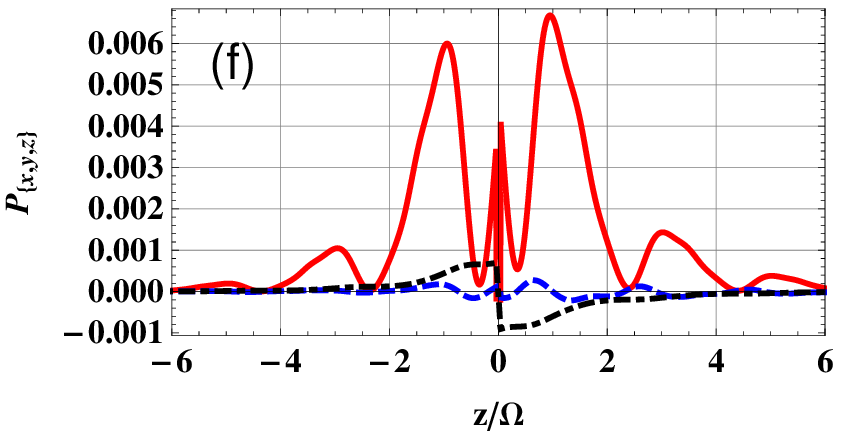}
\end{array}$
\caption{Variation of the Cartesian components of  ${\bf P}(x,z)$ along the $z$ axis when $x=0$, $L_\pm=\pm7.5~\rm{nm}$, and $\gamma^-=\gamma^++45^\circ$.  (a-c) $\gamma^+=25$,  and (d-f) $\gamma^+=150^\circ$. (a) $\kappa/{\ko}=2.6423+ i0.1865$, (b) $\kappa/{\ko}=2.08764 + i0.009246$, (c) $\kappa/{\ko}=1.9159 +i0.009486$, (d) $\kappa/{\ko}=2.6398 + i0.1847$, (e) $\kappa/{\ko}=2.09389 + i0.01000$, and (f) $\kappa/{\ko}=1.9108 + i0.02442$. }
\label{field45L7.5}
\end{center}
\end{figure}

\begin{figure}[!ht]
\begin{center}$
\begin{array}{cc}
\includegraphics[width=2.25in]{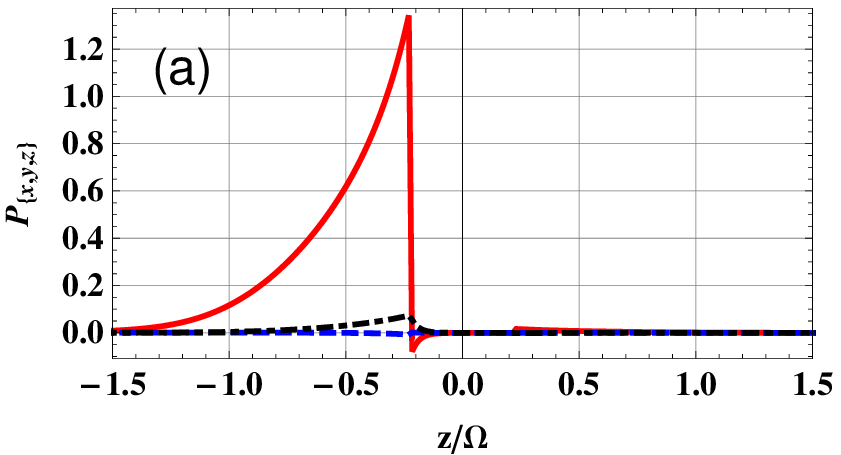}  \includegraphics[width=2.25in]{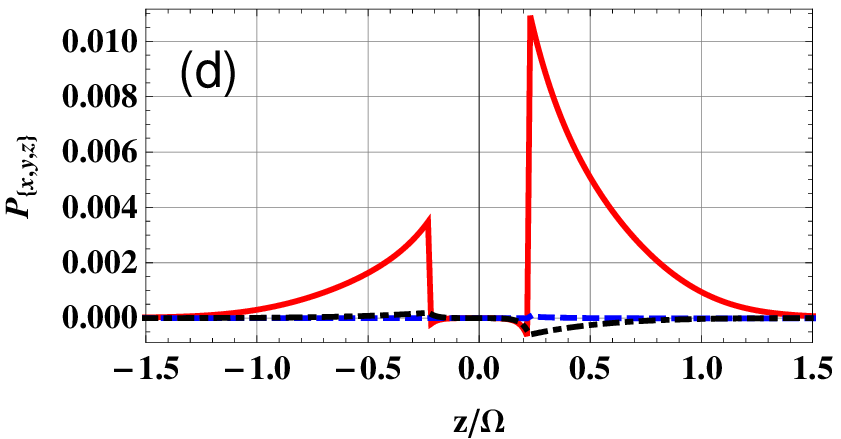}\\
\includegraphics[width=2.25in]{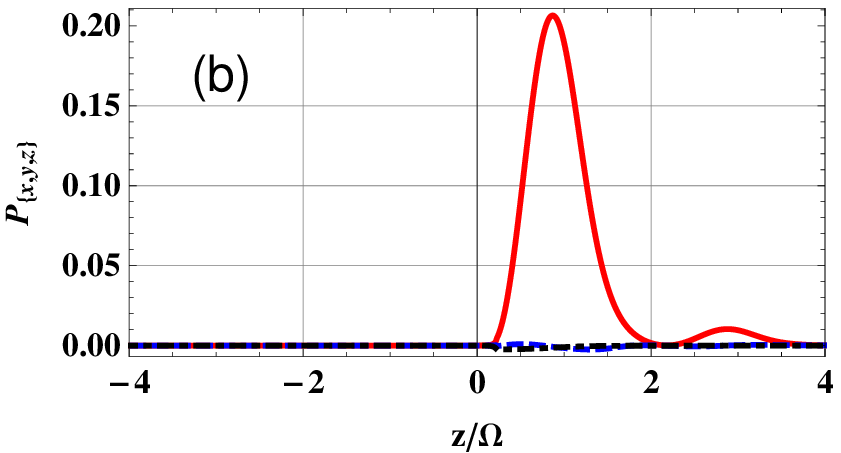} \includegraphics[width=2.25in]{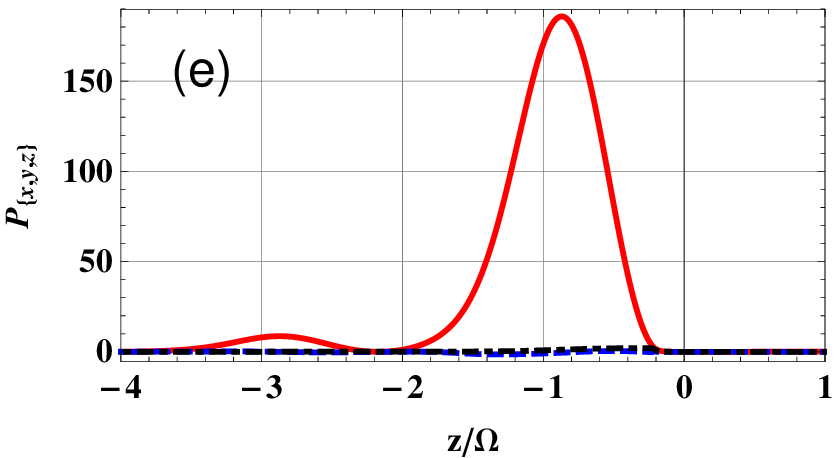}\\
\includegraphics[width=2.25in]{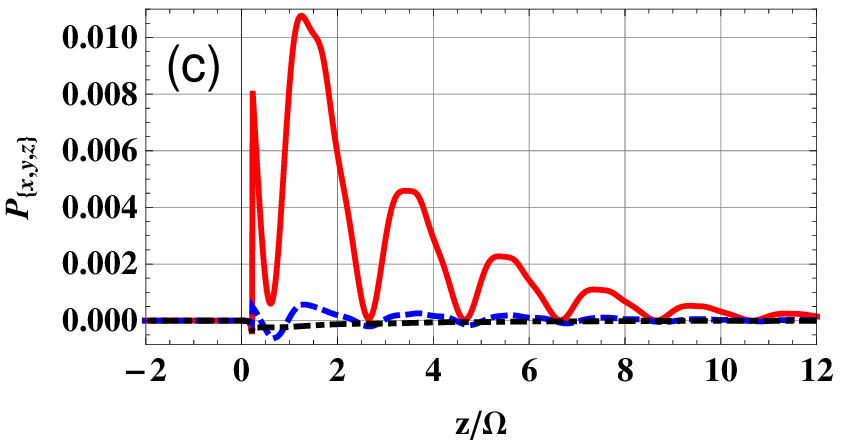} \includegraphics[width=2.25in]{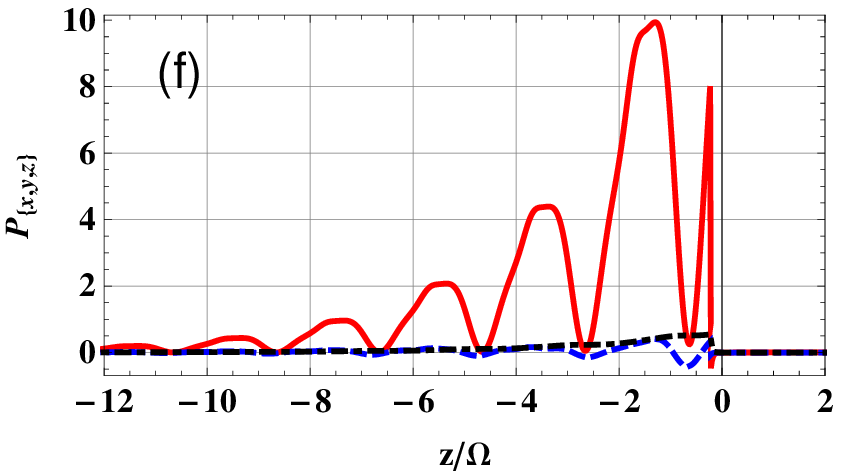} 
\end{array}$
\caption{Same as Fig.~\ref{field45L7.5} except that $L_\pm=\pm45~\rm{nm}$. (a) $\kappa/{\ko}=2.4586 +i0.04246$, (b) $\kappa/{\ko}=2.07725 + i0.003595$, (c) $\kappa/{\ko}=1.8891 + i0.002576$, (d) $\kappa/{\ko}=2.4559 +i0.04227$, (e) $\kappa/{\ko}=2.07916 + i0.003560$, and (f) $\kappa/{\ko}=1.8737 + i0.006143$. }
\label{field45L45}
\end{center}
\end{figure}

\subsection{$\gamma^-=\gamma^++45^\circ$}\label{nrd45}

The solutions of the dispersion equation~(\ref{eq:SPPdisp}) for $\gamma^-=\gamma^++45^\circ$ are given in Fig.~\ref{kappa45} for $\gamma\in \left[0^\circ, 180^\circ\right]$. By virtue of symmetry, the solutions for $\gamma^++180^\circ$ are the same as for $\gamma^+$. We found that five solutions exist for 
the entire range of $\gamma^+$ for $L_\pm=\pm7.5~\rm{nm}$. When $L_\pm=\pm12.5~\rm{nm}$,    five solutions 
exist for $\gamma\in \left[0^\circ, 37^\circ\right]\cup\left[98^\circ, 180^\circ\right]$ but only four for
$\gamma\in \left(37^\circ, 98^\circ\right)$. Further thickening of the metal
to 50~nm  ($L_\pm=\pm25~\rm{nm}$) yields five solutions for
$\gamma\in \left[0^\circ, 36^\circ\right]\cup\left[99^\circ, 180^\circ\right]$ but four for
$\gamma\in \left(36^\circ, 99^\circ\right)$. 
Decoupling of the two metal/SNTF interfaces becomes
very pronounced  for $L_\pm=\pm45~\rm{nm}$, when all solutions found are the same for either (i) a
metal/SNTF interface for which the direction of propagation in the $xy$ plane makes an angel $\gamma^+$ with the morphologically significant plane, or (ii) a
metal/SNTF interface for which the direction of propagation in the $xy$ plane is inclined at
$\gamma^++45^\circ$ to the morphologically significant plane.

Similar to Secs.~\ref{nrd0} and \ref{nrd90}, the solutions can be grouped into three sets, with the same criterions given in Sec.~\ref{nrd0}. Representative field profiles for $15$-nm- and $90$-nm-thick metal slabs are given in Figs.~\ref{field45L7.5} and \ref{field45L45}, respectively.  We selected one value of $\kappa$ from each set at $\gamma^+=25^\circ~{\rm and}~150^\circ$. Whereas $\gamma^+=25^\circ$ corresponds to the propagation in the
$xy$ plane at an angle of $25^\circ$ with respect to the morphologically significant plane in the region
 $z>L_+$ and at $70^\circ$ in the region $z<L_-$, the analogous angles are $30^\circ$ and $15^\circ$, respectively,
 when 
$\gamma^+=150^\circ$. Fig.~\ref{field45L7.5} shows that SPP waves are guided by both interfaces
of thin metal slab, but the power density profile is asymmetric due to the fact that morphologically 
significant planes are not parallel to each other in the two regions occupied by the SNTF. Figure~\ref{field45L45} shows that when $\lmet=90$~nm, any SPP wave propagates predominantly guided by one of the two metal/SNTF interfaces. We deduce from these two figures  that the conclusions drawn in Secs.~\ref{nrd0} and \ref{nrd90} hold true for this case as well, and therefore are general enough.
 
\section{Concluding Remarks}\label{conc}
We formulated and solved a canonical boundary-value problem to examine the characteristics of surface-plasmon-polariton waves guided by a thin metal
slab inserted in a periodically nonhomogeneous sculptured nematic thin film. The morphologically significant planes   on the two
sides of the metal slab could be either parallel to   or twisted with respect to each other. When the 
metal slab is very thin, its two
interfaces with the SNTF couple to each other, thereby generating more modes of SPP-wave propagation. As the metal slab thickness,
the two interfaces decouple. Finally, the two interfaces of a sufficiently thick metal slab independently guide SPP waves.

\noindent {\bf Acknowledgments.}
MF thanks the Trustees of the Pennsylvania State University for a University Graduate Fellowship. AL
is grateful to  Charles Godfrey Binder Endowment
at the Pennsylvania State University for partial support of this work.

 \end{document}